\documentstyle[12pt]{article}

\newcommand{\be}{\begin{equation}}
\newcommand{\ee}{\end{equation}}
\newcommand{\bea}{\begin{eqnarray}}
\newcommand{\eea}{\end{eqnarray}}

\newcommand{\prd}{Phys. Rev. \underline}

\newcommand{\ra}{\rightarrow}
\def\glgt{$\Gamma_L/\Gamma_{tot}$ }
\def\gs {$\Gamma(B\to K^\ast \psi)/\Gamma(B\to K \psi)$ }

\def\beq{\begin{equation}}
\def\eeq{\end{equation}}
\def\noi{\noindent}
\def\bea{\begin{eqnarray}}
\def\eea{\end{eqnarray}}
\def\bei{\begin{itemize}}
\def\eei{\end{itemize}}
\def\bpsi{$B\ra K^{(\ast)} \psi$ }
\def\dk{$D\ra K^{(\ast)}l \nu$ }
\def\hh {heavy-to-heavy }
\def\hhs {heavy-to-heavy scaling }
\def\hl {heavy-to-light }
\def\hls {heavy-to-light scaling }
\def\qq {$q^2$ }
\def\qmax {$q^2_{max}$ }
\def\qqz {$q^2=0$ }

\def\fourthrm{\font\fourthrm=cmr12 at 14pt}
\def\ved{\vec q\, ^2}
\begin{document}
\begin{flushright}
DAPNIA/SPP/94-24,
LPTHE-Orsay 94/15, hep-ph/9408215
\end{flushright}
\vskip 5 mm
\begin{center}
{\bf CRITICAL ANALYSIS OF THEORETICAL ESTIMATES \par
\bf FOR $B$ TO LIGHT MESON FORM FACTORS AND \par
{\bf THE $B \to \psi K(K^{\ast})$ DATA}} \par
\vskip 5 mm
{\bf R. Aleksan} \par
{Centre d'Etudes Nucl\'eaires de Saclay, DPhPE, 91191 Gif-sur-Yvette, France}
\par
\bigskip
{\bf A. Le Yaouanc, L. Oliver, O. P\`ene and J.-C. Raynal} \par
{Laboratoire de Physique Th\'eorique et Hautes Energies\footnote{Laboratoire
associ\'e au
Centre National de la Recherche Scientifique - URA 63}}\\
{Universit\'e de Paris XI, B\^atiment 211, 91405 Orsay Cedex, France}\\
\end{center}
\vskip 5 mm
\noindent {\bf Abstract}\\

\baselineskip 20pt
We point out that current estimates of form factors fail to explain the
non-leptonic decays $B \to \psi K(K^{\ast})$ and that the combination of data
on the semi-leptonic decays $D \to K(K^{\ast})\ell \nu$ and
on the non-leptonic decays $B \to \psi K(K^{\ast})$ (in particular recent
po\-la\-ri\-za\-tion data)
severely constrain the form (normalization and $q^2$ dependence) of the
heavy-to-light meson form
factors, if we assume the factorization hypothesis for the latter. From a
simultaneous fit to \bpsi and \dk data we find that
strict heavy quark limit scaling laws do not hold when going from $D$ to $B$
and must have large
corrections that make softer the dependence on the masses. We find that
$A_1(q^2)$ should increase slower with \qq than $A_2, V, f_+$.
 We propose a simple parametrization of these corrections based on a quark
model or on an extension of the \hhs laws to the \hl case, complemented with an
approximately constant $A_1(q^2)$. We
analyze in the light of these data and theoretical input various theoretical
approaches (lattice
calculations, QCD sum rules, quark models) and point out the origin of the
difficulties encountered by  most of these schemes. In particular we check  the
compatibility of several quark models with the heavy quark scaling relations.\\

\newpage
\pagestyle{plain}
\baselineskip 20pt
\section{Introduction.}
\label{sec-int}

	Heavy-to-heavy meson form factors like $B \to D^{(\ast)}$ obey a very
constraining principle, namely the heavy quark symmetry that relates all form
factors to the Isgur-Wise function \cite{isgur}. However, even in this case,
large corrections of order $1/m_c$ can occur and many uncertainties remain
\cite{nr}, and most importantly the scaling function $\xi$ remains unknown.
	In the case of heavy-to-light meson form factors like $D \to K^{(\ast)}$ or $B
\to K^{(\ast)}$ there are also rigorous results in the asymptotic heavy quark
limit, but which are much weaker, namely relations between the form factors $D
\to K^{(\ast)}$ and $B \to K^{(\ast)}$ at  fixed $\vec q $ near the zero recoil
point $\vec q = 0$, i.e.  $q^2=q^2_{max}$. This is a small kinematical region,
and furthermore, no relation is obtained between the various form factors of a
given hadron decay, unlike the \hh case.
	From semi-leptonic $D \to K^{(\ast)}l\nu$ decays \cite{witherell} we have data
for the form factors,
although with large errors, in a completely different
kinematic region, namely at small $q^2$, and we cannot from these data extract
information on the $B \to K^{(\ast)}$ form factors without knowledge of the
$q^2$ dependence, for which we do not have any rigorous result in the
heavy-to-light case.
In both cases, one must unavoidably appeal to models, like the quark model, or
theory, or just make phenomenological Ans\"atze, for example assume a pole or a
dipole \qq dependence.

	On the other hand, we have very interesting data for the decays $B \to \psi
K^{(\ast)}$, in particular recent polarization data. If we assume
factorization, these data can give us precious informations on the $B \to
K^{(\ast)}$ form factors at a different kinematic point ($q^2=m^2_\psi$)
than the data on semi-leptonic D decays (mainly at $q^2=0$) or the heavy quark
limit QCD scaling laws ($q^2=q^2_{max}$).
	The combination of these D semi-leptonic and B non-leptonic data plus the
rigorous scaling law in the asymptotic heavy quark limit can severely
constrain  the corrections to asymptotic scaling laws and provide rich
information on the
gross features of the $q^2$ dependence of the form factors, as we will see
below. This
is the main object of this paper.

	Present models of non-leptonic B decays are in trouble to
describe the  $B \to \psi K^{(\ast)}$ decays, namely the ratio of decay rates
$\psi K/\psi
K^\ast$ and the $\psi K^\ast$ polarization data simultaneously.
	Moreover, the scaling law in the heavy quark limit is not always verified by
current models of heavy-to-light form factors, and it is important to consider
this matter to gauge the theoretical consistency of models and not only their
phenomenological description of data.

	This paper is organized as follows. In section \ref{sec-bpsi} we address the
simplest question, namely the comparison of the different models of
non-leptonic B decays with the data on $B \to \psi K^{(\ast)}$, to see that
there is a serious difficulty. In section \ref{sec-class} we discuss the
theoretical constraints for the heavy-to-light form factors at the light of the
data to set a general Ansatz for the form factors. We use  these results in
section \ref{sec-disc} where we compare and discuss the different
theoretical schemes. In particular, we make the distinction between quark
models with ad hoc $q^2$ dependence and quark models which derive this
dependence from the wave function overlap. We discuss also the QCD sum rules
and the lattice
QCD results. In subsection \ref{sec-oqm} we propose a quark model that fulfills
the theoretical and some phenomenological requirements stated in
section \ref{sec-class}. Finally, our conclusions are given in section
\ref{sec-conc}. A short overview
of this work has already been given at the Beauty 94 conference
\cite{beauty-94}

\section{$B\to \psi K^{(\ast)}$ data are hardly compatible with current
estimates.}
\label{sec-bpsi}

\par
	To be definite, let us write the form factors :

\[< P_f|V_{\mu}|P_i> = \left ( p_{\mu}^f + p_{\mu}^j - {m_i^2 - m_f^2 \over
q^2}  q_{\mu} \right
) f_+(q^2) + {m^2_i - m^2_f \over q^2}  q_{\mu} f_0(q^2)\]
\[< V_f|A_{\mu}|P_i > = \left ( m_f + m_i \right ) A_1(q^2)
\left ( \varepsilon_{\mu}^{\ast} - {\varepsilon^{\ast}.q \over q^2} q_{\mu}
\right )
\]
\[- A_2(q^2) {\varepsilon^{\ast}.q \over m_f + m_i}
\left ( p^i_{\mu} + p^f_{\mu} -
{m^2_i - m^2_f \over q^2}  q_{\mu} \right ) + 2 m_f
\ A_0(q^2)  {\varepsilon^{\ast} . q \over q^2} q_{\mu}\]
\beq
< V_f|V_{\mu}|P_i > = i {2 \ V(q^2) \over m_f + m_i}  \varepsilon_{\mu \nu \rho
\sigma} p_i^{\nu} \  p_f^{\rho} \varepsilon^{\ast \sigma}.\label{definite}
\eeq
where we use the convention $\epsilon^{0123}=1$.

The point that we want to emphasize in this paper is that the
non-leptonic
decays $B \to \psi K, \psi K^{\ast}$ can help to get hints about two
questions concerning these form factors,
namely: i) the sign and size of the $1/m_Q$ corrections to the asymptotic \hl
scaling relations as well as ii) the gross features of the $q^2$
dependence of the form factors. Of course, we must assume factorization
to relate these decays to the form factors.

In the standard Shifman, Vainshtein and Zakharov (SVZ) \cite{svz} factorization
assumption, that we will call standard SVZ factorization, one deduces the non
leptonic amplitudes from form factors and annihilation constants. There are two
types of two-body decays corresponding to the two different color topologies,
the
so-called classes I and II of Bauer, Stech and Wirbel (BSW) \cite{bsw1},
\cite{bsw2},
respectively proportional to the effective color factors

\beq
a_1 = {1 \over 2} \left [ c_+ \left ( 1 + {1 \over N_c} \right ) + c_- \left (
1 - {1 \over N_c}
\right ) \right ] \qquad a_2 = {1 \over 2} \left [ c_+ \left ( {1 \over N_c} +
1 \right )  +
c_- \left ( {1 \over N_c} -1 \right ) \right ].\label{a1a2}
\eeq
where $c_\pm$ are QCD short distance factors.

The decays we are interested in here, $B \to \psi K, \psi K^\ast$, are of class
II. This standard SVZ factorization, that applies literally with expression
(\ref{a1a2}) using $N_c=3$, is known to fail definitely in class II
decays.
On the other hand there is a distinct
{\it phenomenological factorization} prescription proposed by BSW which
derives  $a_1$ and $a_2$ {\it by fitting} the observed $B_d$
decays. We call this factorization prescription phenomenological in the sense
that $a_1$ and $a_2$ are fitted from the data and not obtained through
theoretical
relations (2). It must be stressed that these fitted coefficients have no
intrinsic meaning in the sense that they are depending on the models used to
estimate the form factors and annihilation constants. The model used has been
traditionally chosen to be the BSW model, later modified by Neubert, Rieckert,
Stech and Xu \cite{bsw2}. These authors  found, from a fit to the two-body $B$
decays:

\beq
|a_1| = 1.11		\qquad |a_2| = 0.21	\ \ \ .
\eeq
 The magnitude of $|a_2|$ is incompatible with the expectation from (2) and the
short distance QCD factors for $N_c=3$: $a_2 \sim 0.1$. More recently, the sign
of $a_2/a_1$ has been unambiguous found positive by considering class III
decays \cite{cleo-94} that depend on the interference between $a_1$ and $a_2$.
 This sign is inconsistent with the once proposed prescription \cite{bsw1} of
taking the
limit $N_c\to \infty$ in eq. (\ref{a1a2}) since  one has $c_+<c_-$ for the
short distance QCD
factors $c_+$, $c_-$.

	 We obtain, within the factorization assumption, the following amplitudes in
the B meson rest frame :

\beq
A \left ( \bar{B}_d^0 \to \psi K \right ) = - {G \over \sqrt{2}} V_{cb}
V_{cs}^{\ast} \ 2 \ f_{\psi} \
m_B \ f_+(m^2_{\psi} ) a_2p
 \eeq

\[A^{pv} \left (\bar{B}^O_d \to \psi (\lambda = 0) K^{\ast}(\lambda = 0) \right
) = - {G
\over \sqrt{2}}  V_{cb} V_{cs}^{\ast} m_{\psi} f_{\psi}\]
\beq
\left [ \left ( m_B + m_{K^{\ast}} \right ) \left (
{p^2 + E_{K^{\ast}} E_{\psi} \over m_{K^{\ast}} m_{\psi}} \right )
A_1(m^2_{\psi}) - {m^2_B \over m_B + m_{K^{\ast}}}
{2p^2 \over m_{K^{\ast}} m_{\psi}} A_2(m^2_{\psi}) \right ] a_2\label{long}
\eeq
\beq
A^{pv} \left ( \bar{B}^0_d \to \psi(\lambda = \pm 1) K^{\ast}(\lambda =
\pm 1) \right ) = - {G \over \sqrt{2}} V_{cb}V_{cs}^{\ast} m_{\psi}
f_{\psi} \left ( m_B + m_{K^{\ast}} \right ) A_1(m^2_{\psi}) a_2
\eeq
\beq
A^{pc} \left ( \bar{B}^0_d \to \psi(\lambda = \pm 1) K^{\ast}(\lambda =
\pm 1) \right ) = \pm {G \over\sqrt{2}}
V_{cb}V_{cs}^{\ast} m_{\psi} f_{\psi}  {m_B \over m_B + m_{K^{\ast}}} 2
V(m^2_{\psi}) a_2 p \ \ .
\eeq

We see that the non-leptonic data plus the factorization hypothesis can give us
information on the form factors at a different kinematic point ($q^2 =
m^2_{\psi})$ than the data on
semi-leptonic $D$ decays (small $q^2$) or the heavy quark limit QCD scaling
laws (at
$q^2_{max}$).

The data for the total rates \cite{cleo-94} are:

\[
BR \left ( \bar{B}_d^0 \to \psi K^0 \right ) = (7.5 \pm 2.4 \pm 0.8) \times
10^{-4}
\]

\[
BR \left ( B_d^0 \to \psi K^{\ast 0} \right ) = (16.9 \pm 3.1 \pm 1.8) \times
10^{-4}
\]
\[
BR \left ( {B}^- \to \psi K^- \right ) = (11.0 \pm 1.5 \pm 0.9) \times
10^{-4}
\]
\[
BR \left ( B^- \to \psi K^{\ast -} \right ) = (17.8 \pm 5.1 \pm 2.3) \times
10^{-4}
\]
\noi and  the recent results of ARGUS \cite{cleo2}, CLEO \cite{cleo2} and
CDF\cite{cdf} concerning the $K^\ast$ polarization in the
$\bar{B}_d \to \psi K^{\ast 0}$ decay, are:

\bea
\Gamma_L /\Gamma_{tot} & > & 0.78 \ (95 \  \% \ C.L.)		\quad \mbox{ARGUS}\,
\nonumber \\
\Gamma_L /\Gamma_{tot} & = & 0.80 \pm 0.08 \pm 0.05 	\quad \mbox{CLEO}
\nonumber \\
\Gamma_L /\Gamma_{tot} & = & 0.66 \pm 0.10 ^{+0.10}_{-0.08} 	\quad \mbox{CDF}
\label{rlexp}\eea

\noindent where $\Gamma_L$ is the partial width for the longitudinal
polarization whose amplitude is
given by (\ref{long}). \par

	As we have pointed out these decays are affected by
the phenomenological factor $a_2$ which is not well known from other sources.
To avoid this uncertainty, we will consider the
ratio of the total rates

\beq
R \equiv {\Gamma \left ( \bar{B}_d^0 \to \psi K^{\ast 0} \right ) \over \Gamma
\left (
\bar{B}_d^0 \to \psi K^{0} \right )} = 1.64 \pm 0.34 \quad\mbox{CLEO II}\,\,
\cite{browder} \label{rstar}\eeq

\noi and the polarization ratio for $\psi K^{\ast 0}$~:

\beq
R_L \equiv {\Gamma_L \left ( \bar{B}_d^0 \to \psi K^{\ast 0} \right ) \over
\Gamma_{tot} \left ( \bar{B}_d^0 \to \psi K^{\ast 0} \right )}
\label{rl}\eeq
that are independent of $a_2$.

	Assuming factorization, any model or Ansatz on the heavy-to-light meson form
factors will give a prediction for these ratios that can be compared to
experiment. These  ratios in terms of the form factors
are given by eqs. (\ref{rlform}) and  (\ref{r}).

{}From these formulae one can already conclude qualitatively that :

i) To get $R_L$ sufficiently large, one needs $V/A_1$ and $A_2/A_1$ to be small
enough.

ii) To get $R$ not too large $f_+/A_1$ must not be too small.

 We will consider the predictions for these ratios from the following
theoretical schemes :
\bei
\item	{1.} Pole model of Bauer, Stech and Wirbel (BSWI) \cite{bsw1}.
\item	{2.} Pole-dipole model of Neubert et al. (BSWII) \cite{bsw2}.
\item	{3.} Quark model of Isgur, Scora, Grinstein and Wise (ISGW) \cite{isgw}.
\item	{4.} QCD sum rules (QCDSR) \cite{ball}.
\eei
	The results are given in Table \ref{tab-RRL}. We do not include here the
lattice results on the form factors \cite{abada} because they are still
affected by
large errors; we will discuss these results in section \ref{sec-disc}. The
conclusion of the table is that there is a problem for all known theoretical
schemes since both ratios $R$ and $R_L$ cannot be described at the same time. A
priori there are three possible explanations :

i) The theoretical schemes for {\it form factors} are to be blamed for the
failure.

ii) The experimental numbers are not to be trusted too  much.

iii) The basic BSW factorization assumption, which allows to relate \bpsi to
the form factors, is wrong  for class II decays.

In section \ref{sec-class}, we will explore the first possibility by
trying to formulate form factors satisfying the relevant theoretical principles
and being able to describe the experimental situation. In the light of our
discussion
in the latter section  we will return, in section \ref{sec-disc}, to these
models and try an analysis of their theoretical difficulties.

%___________________________________________________________________________
\begin{table}
\centering
\begin{tabular} {|c|c|c|c|c|c|}
\hline
 &  $\frac{\Gamma(K^\ast)}{\Gamma(K)}$ & $\frac{\Gamma_L}{\Gamma_{tot}}$ &
$\frac {A_2^{sb}(m_\psi^2)}{A_1^{sb}(m_\psi^2)}$&$
 \frac {V^{sb}(m_\psi^2)}{A_1^{sb}(m_\psi^2)}$&$ \frac
{f_+^{sb}(m_\psi^2)}{A_1^{sb}(m_\psi^2)}$\\ \hline
BSWI \cite{bsw1}&4.23	&0.57 & 1.01& 1.20 &1.23\\ \hline
BSWII \cite{bsw2}&1.61	&0.36 &1.41  & 1.77 &1.82\\ \hline
ISGW \cite{isgw}& 1.71&0.07 & 2.00 & 2.58 &2.30\\ \hline
QCDSR \cite{ball}& 7.60&0.36 & 1.19 & 2.66  & 1.77\\ \hline
Soft-Pole (CDF) & 2.15 & 0.45 & 1.08 & 2.16 & 1.86 \\ \hline
CLEO II \cite{cleo2}, \cite{browder} 	&    $1.64\pm 0.34$   &	    $0.8\pm0.1$ &
& &\\
CDF \cite{cdf}	&    -   &	    $0.66\pm0.1$ & & &\\\hline
\end{tabular}
\caption{\it Comparison of different models, a QCD Sum Rules calculation and
our prefered Ansatz (Soft-Pole as defined in table 3) to experiment. In the
fifth line CDF means fit to this data as explained in table 3.}
\label{tab-RRL}
\end{table}

%*******************************************

\section{Phenomenological \hls form factors confronted to $B$ and $D$
experiments.}
\label{sec-class}

\subsection{Setting  the problem.}
\label{sec-setting}
	Our aim will be to perform a combined  experimental and theoretical study of
form factors, simultaneously for both of \dk and \bpsi decays, assuming BSW
factorization for the latter. We would like to proceed as much independently
as possible
of the detailed theoretical approaches, using {\it general Ans\"atze
that respect the \hl asymptotic scaling laws}, some of them being complemented
by ideas derived from \hhs law formulae. Only guided by  rigorous theoretical
laws and some commonly admitted theoretical prejudices, we will try to display
general trends suggested by the experiment. Finally we will underline that
experiment, as
it stays today, is not easy to account for in a theoretically reasonable
manner. We will also advocate the use of a Quark Model inspired prescription,
that we call the ``QMI'' Ansatz, an extension of some \hhs relations to the \hl
system. Although not fully successful, this model is able to account roughly
for a large set of data.

	First, let us review available data. Besides  the indirect indications
coming from the above $B \to \psi K(K^{\ast})$ non leptonic data complemented
by
the BSW factorization assumption, there are data on the $D \to K(K^{\ast}) \ell
\nu$ form factors, mainly around $q^2 = 0$. We shall use the world average
\cite{witherell}:

\begin{eqnarray}
f^{sc}_+(0) & = & 0.77 \pm 0.08  \nonumber \\
V^{sc}(0) & = & 1.16 \pm 0.16  \nonumber \\
A^{sc}_1(0) & = & 0.61 \pm 0.05  \qquad  \nonumber \\
A^{sc}_2(0) & = & 0.45 \pm 0.09
\label{dexp}\end{eqnarray}

\bea
V^{sc}(0)/A^{sc}_1(0) = 1.9 \pm 0.25 \nonumber \\
A^{sc}_2(0)/A^{sc}_1(0) = 0.74 \pm 0.15  \ \ \ .
\label{drexp}\eea

As to the $q^2$ dependence the indications are poor except for the $f_+$ form
factor where good indications seem to support the relevant vector meson pole
dominance. We will use these indications for the \qq dependence only in a
second
stage.

To organize the discussion which must handle a great number of possibilities,
we will often first concentrate on the evolution from $D$ to $B$ of the {\it
ratios} between different form factors ($A_2/A_1$, $V/A_1$, etc.) for which we
can formulate more general statements, and then consider the values  and
evolutions of form factors themselves ($A_1$, $f_+$, etc.) which involve
additional assumptions. The advantage of discussing first the ratios is that we
can then draw more direct conclusions from the $B \to \psi K(K^{\ast})$   data
which depend only on the ratios of form factors and already exclude a number of
possibilities before considering absolute branching ratios which also involve
the unknown $a_2$ defined in eq. (\ref{a1a2}).

%------------------------------------------
\subsection{Asymptotic scaling laws for the \hl form factors.}
\label{sec-asym}

	What can be learned from the theory ? The only exact results take the form of
asymptotic theorems \cite{isgurhl} valid for the initial quark mass $m_Q$
large with respect to a typical scale $\Lambda$, of QCD,  to the
final meson mass, $m_f$, and to the final momentum, $\vec q$ \footnote{Unless
specified otherwise, we use the initial
meson rest frame.}:

\[ \hat f_+(\ved),\quad \hat V(\ved),\quad \hat A_2(\ved) = {m_Q}^{\frac 1 2}
\left(1+O\left(\frac \Lambda{m_Q}\right)+
O\left(\frac {|\vec q|}{m_Q}\right)+O\left(\frac {m_f}{m_Q}\right)\right)\]

\be \hat A_1(\ved) = \left({m_Q}\right)^{-\frac 12} \left(1+O\left(\frac
\Lambda{m_Q}\right)+
O\left(\frac {|\vec q|}{m_Q}\right)+O\left(\frac
{m_f}{m_Q}\right)\right)\label{hl}\ee
where we have used ``hats'' on form factors to indicate that they depend {\it
three-momentum, the natural variable in the \hl case}:
\beq
\hat f(\ved) = f(q^2), \qquad \mbox{with}\qquad \ved = \left(\frac
{m_i^2+m_f^2-q^2}{2 m_i}\right)^2-m_f^2\label{hats}
\eeq

The asymptotic scaling law (\ref{hl}) allows to  relate the
form factors, say $D \to K$ and $B \to K$, at small recoil
$|\vec{q}| \ll m_D$ (i.e. close to  $q^2_{max}$ for each process):

\[
{\hat f^{sb}_+(\ved) \over \hat f^{sc}_+(\ved)},\quad {\hat V^{sb}(\ved) \over
\hat V^{sc}(\ved)},\quad
{\hat A^{sb}_2(\ved) \over\hat A^{sc}_2(\ved)} =
\left ( {m_B \over m_D} \right )^{{1 \over 2}} \left(1+O\left(\frac
\Lambda{m_D}\right)+
O\left(\frac {|\vec q|}{m_D}\right)+O\left(\frac {m_f}{m_D}\right)\right)\]

\beq
{\hat A^{sb}_1(\ved) \over \hat A^{sc}_1(\ved)} = \left ( {m_D \over m_B}
\right )^{{1 \over
2}} \left(1+O\left(\frac \Lambda{m_D}\right)+
O\left(\frac {|\vec q|}{m_D}\right)+O\left(\frac
{m_f}{m_D}\right)\right),\label{hlr}\ \ \
\eeq
which hold for $m_B$ and $m_D$ much larger than $\Lambda$, the spectator quark
and final meson masses as well as the final meson momentum.

 \subsection{Failure of the simple-minded extrapolation from $D$ to $B$
according to the asymptotic scaling law.}
\label{sec-failu}

In this subsection we will stress qualitatively that $B\ra K^{(\ast)}\psi$ data
seem to exclude a simple-minded extrapolation from $D\ra K^{(\ast)} l\nu$ data
at $q^2=0$ according to the
\hl asymptotic scaling law. Since the momentum $\vec q$ is different in the two
above-mentioned sets of data, an hypothesis on the $q^2$ dependence is needed.
However, before coming to more quantitative discussions (see \ref{sec-confr}),
we will simply assume that this dependence does not change the qualitative
consequences of the scaling laws.

The ratio \glgt is given by:

\be \frac{\Gamma_L(B\to K^\ast \psi)}{\Gamma_{tot}(B\to K^\ast \psi)}=\frac
{\left(3.162 -1.306 \frac {A_2^{sb}(m_\psi^2)}{A_1^{sb}(m_\psi^2)}
\right)^2}{2\left[1+0.189\left(\frac{V^{sb}(m_\psi^2)}
{A_1^{sb}(m_\psi^2)}\right)^2
\right] +\left(3.162 -1.306 \frac {A_2^{sb}(m_\psi^2)}{A_1^{sb}(m_\psi^2)}
\right)^2}\label{rlform}\ee

{}From this expression it is apparent that $A_2/A_1$ must not be too
large\footnote{Strictly speaking very large values, $A_2/A_1\ge 3.9$,  could
also
account for a large $R_L$, but these are unrealistic.} in view of the large
experimental value of $R_L$ (\ref{rlexp}), all the more if $V/A_1$ is large.
For example, setting $V=0$ we get the very conservative upper bound $A_2/A_1
\le 1.3$ for $R_L>0.5$. For a more realistic value of $V/A_1\simeq 2$, the
upper bound becomes $A_2/A_1 \le 1$. Now, according to strict application of
the asymptotic scaling laws described above, $A_2/A_1$ ($V/A_1$) would be
multiplied at fixed $\vec q$ by $m_B/m_D= 2.83$. From the central experimental
$D$
value, $A^{sc}_2/A^{sc}_1=0.74$ ($V^{sc}/A^{sc}_1=1.9$), one gets
$A^{sb}_2/A^{sb}_1=2.09$ ($V^{sb}/A^{sb}_1= 5.38$) at $q^2= 16.56$ GeV
(corresponding in $B$ decay to the same $\ved$ as $q^2=0$ in $D$ decay). This
is in
drastic contradiction with experiment unless there is an  unexpectedly strong
$q^2$ variation down to $q^2=m_{\psi}^2$. A naive insertion of these values in
eq. (\ref{rlform}) would indeed give $R_L=0.014$ which is 4 to 5 sigmas away
from the most favorable CDF value. Clearly the message is that {\it a softening
of the increase  with respect to the asymptotic scaling law is required}.

%-----------------------------------------
\subsection{Extending the \hh Isgur-Wise scaling laws into a heavy-to-light
class of Ans\"atze.}
\label{sec-ext}

At this point it is useful to notice that there is an overlap between the
domains of validity of the \hh  and  \hl scaling laws, namely when \hhs laws
are applied when both masses are large but the final quark mass is sensibly
smaller than the initial one:
$m_{Q_f}\ll  m_{Q_i}$.
In this domain the \hhs law provides corrections to the asymptotic \hls law of
order $m_{Q_f}/m_{Q_i}$ that go in the desired direction of a softening. This
will be explained in the next subsection.

\subsubsection{Reminder about asymptotic scaling laws for \hh transitions.}

It is well known that a much stronger set of relations than the one in
subsection \ref{sec-asym} comes from the Isgur-Wise scaling laws \cite{isgur}
for transition form factors between two heavy quarks. Using the notations in
\cite{nr} :

\[{ \sqrt{4 m_{P_i} m_{P_f}} \over  m_{P_i} + m_{P_f} }  f_+(q^2) =
{ \sqrt{4 m_{P_i} m_{P_f}} \over  m_{P_i} + m_{P_f} } {f_0(q^2) \over  1 -
{q^2 \over \left ( m_{P_i} + m_{P_f} \right )^2}} =
{ \sqrt{4 m_{P_i} m_{V_f}} \over m_{P_i} + m_{V_f} } V(q^2) =\]

\beq
= { \sqrt{4 m_{P_i} m_{V_f}} \over m_{P_i} + m_{V_f} } A_0(q^2) = { \sqrt{4
m_{P_i}
m_{V_f}} \over  m_{P_i} + m_{V_f} } A_2(q^2) = { \sqrt{4 m_{P_i} m_{V_f}}
\over m_{P_i} + m_{V_f} }{A_1(q^2) \over  1 - {q^2 \over \left ( m_{P_i} +
m_{V_f} \right )^2} }  = \xi (v_i.v_f)
\label{hh}\eeq
for $m_{P_i}$,  $m_{P_f}$ and $m_{V_f}$ much larger than the typical scale
$\Lambda$ of QCD. It must be added that in the same limit $m_{P_f}$ and
$m_{V_f}$ are in fact equal so that our writing of  different masses is only
meant for later use in the real subasymptotic regime, where they are very
different ($m_K\ne m_{K^\ast}$).

The denominator that divides $A_1(q^2)$ is a straightforward consequence of the
heavy quark symmetry and of the definition of the different form factors. It
has not the meaning of a dynamical pole related to some intermediate
state\footnote{Although the singularity happens to fall at the branching point
of a t-channel cut.}. It is still in the mathematical sense a pole of the ratio
$A_2(q^2)/A_1(q^2)$ etc, and we shall call it for simplicity the ``kinematical
pole''.

 When eq. (\ref{hh}) may be
applied, it is much stronger than the \hl constraint (\ref{hl}).
 Using this \hh  relation (\ref{hh}) for two different values
($m_{P_i}=m_B, m_D$) we automatically obtain the \hl one (\ref{hl}) when one
makes $m_{P_i}$ much larger than $m_f$\footnote{In our notations $m_f$
represents generically $m_{P_f}$ and $m_{V_f}$.}. Indeed, at fixed $\vec{q}$ ,
$v_i.v_f=\sqrt{1+\ved/m_f^2}$ (in the rest frame of the initial meson) is
fixed. It is also simple to show that :
\beq
{4 m_{P_i}m_f \over \left ( m_{P_i} + m_f \right )^2} {1 \over  1-{q^2
\over \left ( m_{P_i} + m_f \right )^2} }=\frac 2 {1+v_i.v_f}
\label{vvprime}\eeq
so that the preceding equations write finally in terms of masses and the fixed
$\vec q$, with $m_{P_i}\gg m_f$ :

\[{ 2\left(\frac {m_{P_f}}{ m_{P_i}} \right)^{1/2}(1- \frac{m_{P_f}}{m_{P_i}}
+.. ) }  \hat f_+(\ved) =\frac{\sqrt{ m_{P_i} m_{P_f}}} {m_{P_f}+E_{P_f}}(1 +
\frac{m_{P_f}}{m_{P_i}} )
 {\hat f_0(\ved)} =\]

\[2\left(\frac {m_{V_f}} { m_{P_i}}\right)^{1/2}(1- \frac{m_{V_f}}{m_{P_i}} +..
)
 \hat V(q^2) =
 { 2\left(\frac {m_{V_f}} { m_{P_i}}\right)^{1/2}(1- \frac{m_{V_f}}{m_{P_i}}
+.. ) } \hat A_0(\ved) =\]

\beq{ 2\left(\frac {m_{V_f}}{ m_{P_i}} \right)^{1/2}(1- \frac{m_{V_f}}{m_{P_i}}
+.. ) } \hat A_2(\ved) = { \frac{\sqrt{ m_{P_i} m_{V_f}}} {m_{V_f}+E_{V_f}}(1 +
\frac{m_{P_f}}{m_{P_i}} )}{\hat A_1(\ved)}  = \xi(E_f/m_f)
\label{hhl}\eeq
where $E_f=E_{V_f}\simeq E_{P_f}$ are the final energies in the initial rest
frame, $E_f=\sqrt{m_f^2+\ved}$.  The ``hats'' on form factors have been
defined in eq (\ref{hats}), and we have used

\be 1- \frac{q^2}{(m_{P_i}+m_f)^2}= 2\frac{ m_{P_i}(m_f+E_f)}{(m_{P_i}+m_f)^2}
\label{energie}\ee

We see that eq. (\ref{hhl}) includes specific values for the $O(m_f/m_{P_i})$
corrections to the \hl scaling law (\ref{hl}). Of course these
corrections are  in principle only valid if $m_f$ is
heavy. Any specific model claiming to handle the domain of mass $\Lambda\ll
m_f\ll m_{P_i}$ should obviously satisfy the relations (\ref{hhl}).

 An essential effect displayed by formula (\ref{hhl}) is that it softens the
asymptotic scaling relation (\ref{hl}), i.e. it leads to a slower increase
(decrease) of $A_2$, $V$, $f_+$ ($A_1$) when the initial mass $m_{P_i}$
increases
at fixed $\vec q$.

Another very important aspect of the Isgur-Wise relations (\ref{hh}) is that
all the ratios of form factors for the same quark masses and at  the same
transfer  \qq (or equivalently $\vec{q}$) are completely fixed by the theory:
{\it quite strikingly, the form factors must be all asymptotically equal at
$q^2=0$, and therefore their ratios are completely independent of the masses;
this extends to any fixed \qq provided that \qq is small with respect to
$m_B^2$} :

\be f_+(0)=f_0(0)=V(0)=A_0(0)=A_2(0)=A_1(0)=\xi\left((v_i\cdot
v_f)_{q^2=0}\right)\label{zero}\ee
where
$$ (v_i\cdot v_f)_{q^2=0}=\frac {m_{P_i}^2+m_{f}^2}{2 m_{P_i}m_{f} }$$

This is to be contrasted with the \hls (\ref{hl}) which leaves these ratios
{\it at the same \qq }  undetermined, because
the dependence in $\vec{q}$ is undetermined.
Even when non asymptotic effects are included, we expect from eq. (\ref{zero})
the ratios $f_+/A_1$, $V/A_1$ and $A_2/A_1$ to behave very softly as a function
of the heavy initial mass at $q^2=0$.

%*******************************************************************

\subsubsection{The basic Soft-scaling-Pole Ansatz}
\label{sec-basic}

We now formulate our model based on an extension of the \hh scaling relations
(\ref{hh}). Let us first assume that we are in a situation described in the
preceding section with two heavy quarks but $m_{i}\gg m_f\gg \Lambda$. As we
have argued, the form factors obey the \hls relations (\ref{hl}) with specific
form factor
ratios and specific $O(m_f/m_i)$ corrections, eq. (\ref{hhl}). To these, one
should also add the unknown $O(\Lambda/m_f)$ corrections to the heavy quark
symmetry.

Let us now consider the intermediate region where the final quark ceases to be
heavy. Our ignorance comes from the fact that the $O(\Lambda/m_f)$ corrections
become large and may totally modify the above mentioned specific relations. Our
hypothesis will be that it is not so, i.e. that using {\it some} of the
features of eq. (\ref{hh}) is indeed a good approximation. This hypothesis,
although admittedly
arbitrary, may be empirically justified by the fact (see section
\ref{sec-failu}) that data demand a softened \hls, and that formula (\ref{hh})
or equivalently  (\ref{hhl}) does present such a behaviour. Theoretical
arguments in favor of the present Ansatz will come below and in section
\ref{sec-theor}.

 It is obvious that an unrestricted extension of Isgur Wise formulae (\ref{hh})
cannot describe quantitatively the form factors for a simple reason: the \dk
form factors at $q^2=0$, eq. (\ref{dexp}), are obviously not equal to each
other, contrary to what is predicted by eq. (\ref{zero}); and the formula
will also completely fail for \bpsi, since it would predict from formula
(\ref{r}) a much too large
ratio $\Gamma(K^\ast)/\Gamma(K)\simeq 4$. This is after all expected because we
do
not believe that the $D$ is heavy, not to speak about the $K$ or $K^\ast$. But
notwithstanding this failure we will try to apply to the \hl case the \qq and
mass dependence implied by formulae (\ref{hh}). The problem of $D$
decays can be trivially cured by assuming, as we shall do, a ``rescaling" of
each form factor to put it
in agreement with the $D$ data at $q^2=0$. We  assume also that these
rescaling factors ($r_+, r_V, r_1, r_2$) are independent of the initial heavy
quark mass and of $q^2$. In other words, we assume the $O(\Lambda/m_f)$
corrections to be properly taken into account by these constant rescaling
factors. Let us thus start from eq. (\ref{hh}), multiply for conveniency the
l.h.s.
and the rhs by $(1+v_i\cdot v_f)/2$, and rescale the form factors as
mentioned above. We obtain:

\[{  m_{P_i} + m_{P_f}   \over \sqrt{4 m_{P_i} m_{P_f}}}\left[ 1 - {q^2 \over
\left ( m_{P_i} +
m_{P_f} \right )^2}\right]\frac{ f_+(q^2)}{r_+} =
 {  m_{P_i} + m_{V_f}  \over \sqrt{4 m_{P_i} m_{V_f}}}
 \left[ 1 - {q^2 \over \left ( m_{P_i} +
m_{V_f} \right )^2} \right]\frac{V(q^2)}{r_V} =\]

\beq
=  {  m_{P_i} + m_{V_f}   \over \sqrt{4 m_{P_i} m_{V_f}}} \left[ 1 - {q^2 \over
\left ( m_{P_i} +
m_{V_f} \right )^2} \right]\frac{A_2(q^2)}{r_2} = {  m_{P_i} + m_{V_f}   \over
\sqrt{4 m_{P_i} m_{V_f}}} {A_1(q^2) \over r_1}  = \eta(\vec q, m_f)
\label{sp}\eeq
where $m_f$ is as usual the final meson mass: $m_{P_f}$ or $m_{V_f}$. In fact,
to conform with the asymptotic Isgur-Wise \hhs, the rescaling parameters $r_+,
r_V, r_2, r_1= 1+O(\Lambda/m_f)$ should depend on the final active quark mass
$m_{q_f}$ and tend to one when it goes to infinity, but this does not matter
here since the final quark will remain the $s$-quark all over this study.
In formula (\ref{sp}) we have introduced a function $\eta(\vec q, m_f)$, since
$\vec q$ is the natural variable in the \hls case.

It should be repeated that at fixed $\ved$ formula (\ref{sp}) corrects the
asymptotic scaling (\ref{hl}) by the replacement $m_{P_i}^{1/2}\ra
(m_{P_i}+m_f)/m_{P_i}^{1/2}$. This correction to the asymptotic scaling induces
a
softer behaviour than the asymptotic scaling, eq.  (\ref{hl}), i.e. a slower
increase/decrease,
when $m_{P_i}$
increases. Consequently eq
(\ref{hlr}) becomes specified into:

\[
{\hat f^{sb}_+(\ved) \over \hat f^{sc}_+(\ved)},\quad {\hat V^{sb}(\ved) \over
\hat V^{sc}(\ved)},\quad
{\hat A^{sb}_2(\ved) \over\hat A^{sc}_2(\ved)} =
\left(\frac{m_B+m_f}{m_D+m_f}\right)\left ( {m_B \over m_D} \right )^{-{1 \over
2}} \]

\beq
{\hat A^{sb}_1(\ved) \over \hat A^{sc}_1(\ved)} =
\left(\frac{m_D+m_f}{m_B+m_f}\right)\left ( {m_D \over m_B} \right )^{-{1 \over
2}} ,\label{spr}\ \ \
\eeq

Another important consequence of formula (\ref{sp}) is that  at $q^2=0$ the
form factor ratios stay constant for fixed final masses when $m_{P_i}$ varies,
 except for a small variation of  $f_+/A_1$
of the order $O((m_{V_f}-m_{P_f})/m_{P_i})$:

\[ \frac {A_2(0)}{A_1(0)}=\frac {r_2}{r_1},\qquad \frac {V(0)}{A_1(0)}=\frac
{r_V}{r_1}\]
\be \frac {f_+(0)}{A_1(0)}=\frac {r_+}{r_1}\left(\frac {m_{P_f}}{m_{V_f}}
\right)^{\frac 1 2}\frac {m_{P_i}+m_{V_f}}{m_{P_i}+m_{P_f}}\,
\frac {\eta(\vec q_1, m_{P_f})}{\eta(\vec q_2, m_{V_f})}\label{zerohl}\ee
where $|\vec q_1|=(m_{P_i}^2-m_{P_f}^2)/2m_{P_i}$ and $|\vec
q_2|=(m_{P_i}^2-m_{V_f}^2)/2m_{P_i}$ correspond to $q^2=0$ for a pseudoscalar
and a vector final meson respectively.

  The formula (\ref{sp}) is of course a purely phenomenological assumption,
although suggested by our quark model analysis in the weak binding limit (see
below); it still presents a large arbitrariness, corresponding to the freedom
of $\eta(\vec q, m_f)$.

Until section \ref{sec-abs} we shall not need to specify the function
$\eta(\vec
q,m_f)$ as we shall be concerned only with form factor ratios. For completeness
we shall now simply state our choice, referring to  section \ref{sec-abs} for
justifications:

\be \eta(\vec q, m_f)=1\label{one}\ee

\subsubsection{Theoretical justifications}
\label{sec-theor}

\paragraph {a) Quark model:}

How do we justify this Ansatz and in particular this ``rescaling'' procedure ?
We are mainly motivated by the fact that the general structure of the
Isgur-Wise relations (\ref{hh}) will be shown to appear also in the heavy to
light case in a quark model with weak binding treatment described below, the
Orsay Quark Model (OQM). It gives the kinematical pole factor, differentiating
$f_+, A_2, V$ from $A_1$. It also displays the $O(m_f/m_i)$ corrections
predicted by the \hhs laws. On the other hand the quark model analysis leads to
expect two types of  $O(\Lambda/m_f)$ corrections:

i) Corrections  taking into account the finite mass of the spectator quark,
which are present in the weak binding treatment.

ii) Corrections to the weak binding limit, not included in the OQM.

In this model  the dominant correction to asymptotic scaling and the dominant
features of \qq dependence are represented by the Ansatz (\ref{sp}), while
additional corrections
are present but are small.

\paragraph {b) $B \to K^\ast \gamma$:}

An amusing example that exhibits the same trends as we advocate is
provided by the  $B \to K^\ast \gamma$ form factors. Defining the $T_i$ form
factors as follows,

\[
<K^\ast, k,\epsilon | \bar s \sigma^{\mu\nu} q_\nu \frac{1+\gamma_5}2 b|B,
p>=-
2 \epsilon_{\mu\nu\lambda\sigma}\epsilon^{\ast\nu}p^\lambda k^\sigma T_1(q^2)
 -i\bigg[\epsilon^\ast_\mu (m_B^2-m_{K^\ast}^2)-\]
\be \epsilon^\ast\cdot q (p+k)_\mu \bigg]
T_2(q^2)  -i \epsilon^\ast\cdot q\left[ q_\mu-\frac {q^2}{m_B^2-m_{K^\ast}^2}
(p+k)_\mu\right] T_3(q^2)\label{gamma}\ee
it is well known that, for $q^2=0$, using the identity
$\sigma_{\mu\nu}\gamma_5=\frac i 2
\epsilon_{\mu\nu\lambda\sigma}\sigma^{\lambda\sigma}$, one obtains the exact
relation:

\be T_1(0)=T_2(0)\label{egalite}\ee

It has also been shown \cite{iwbsg} that

\[ T_1(q^2)= \sqrt{m_Q}\left(1 + O\left(\frac \Lambda{m_Q}\right)+
O\left(\frac {|\vec q|}{m_Q}\right)\right)\]
\be T_2(q^2)= \frac 1 {\sqrt{m_Q}}\left(1 + O\left(\frac \Lambda{m_Q}\right)+
O\left(\frac {|\vec q|}{m_Q}\right)\right)\label{hlbsg}\ee

In the \hh case one may also show that

\be{ \sqrt{4 m_{P_i} m_{V_f}} \over m_{P_i} + m_{V_f} } T_1(q^2)=
{\sqrt{4 m_{P_i} m_{V_f}} \over m_{P_i} + m_{V_f} } \frac {T_2(q^2)}
{1 - {q^2 \over \left ( m_{P_i} +
m_{V_f} \right )^2}}=\frac 1 2\xi(v\cdot v')\label{hhbsg}\ee
which is of course fully compatible with the relation (\ref{egalite}).

But the new thing here is that {\it the relation (\ref{egalite}) remains exact
when
the final quark becomes light}. Since the scaling behaviours of the $T_1$ and
$T_2$ differ in the vicinity of \qmax, (\ref{hlbsg}), the equality
(\ref{egalite}) is a clear indication that the \qq behaviour of both form
factors differs sensibly.
 For example a pole dominance hypothesis for both form factors is totally
excluded by these relations. Furthermore, an extension of
relation (\ref{hhbsg}) to the \hl domain, as we have suggested in the section
\ref{sec-basic}, would directly comply with both relations (\ref{egalite}) and
(\ref{hlbsg}).  Finally let us  insist that this is by no means a proof of our
Ansatz, it is simply a hint that it may point towards the right direction.

\paragraph {c) Matrix elements:}

Some light can be shed on our Ansatz (\ref{sp}), as far as the mass dependence
is concerned, by noting that it amounts to assume that the matrix elements
satisfy an uncorrected asymptotic scaling. To illustrate this point let us
consider a final vector meson $V_f$ with a polarization $\epsilon^T$ orthogonal
to the initial and final meson momenta.

{}From eqs. (\ref{definite}) and (\ref{sp}) the matrix elements scale as
follows:

\[ \frac{<V_f, \epsilon^T, \vec q |A_\mu|P_i>}{\sqrt{4m_B m_{V_f}}} =r_1
\,\eta(\vec q, m_{V_f}) \epsilon_\mu\]

\be \frac{<V_f, \epsilon^T, \vec q |V_\mu|P_i>}{\sqrt{4m_B m_{V_f}}}= i r_V \,
\frac{\eta(\vec q, m_{V_f})} {1+v^0_f} \,{\left(\vec v_f \times \vec
\epsilon\,^T\right)_\mu}\label{matel}\ee
where $v_f^\mu=p_f^\mu/m_f$ and eq. (\ref{energie}) has been used .

In this example it is clear that the matrix elements scale exactly like
$\sqrt{m_{P_i} m_f}$, which is their asymptotic \hls behaviour. Our claim in
favor of the softened scaling eq (\ref{sp}) is equivalent to the statement that
the matrix element asymptotic scaling laws are not corrected at non-asymptotic
masses.
In other words {\it our softened scaling Ansatz is equivalent to a precocious
asymptotic scaling of the matrix elements}.

\paragraph {d) QCD sum rules, lattice calculations:}

We will argue in section \ref{sec-qcd} that QCD sum rules qualitatively  favor
the \qq dependence of the form factor ratios as depicted in eq. (\ref{sp}),
i.e. they generally show an increase of the ratios $A_2/A_1, V/A_1, f_+/A_1$
with \qq not very different from the increase due to the kinematical pole
$1/(1-q^2/(m_{P_i}+m_f)^2)$.

Lattice calculations \cite{abada} on their side, favor a softened \hls, as a
function of the heavy masses, for the leptonic decay constant $F_P$ and for the
form factors except $A_2$. However, if this result seems strongly established
for the leptonic decay constants, the $m_Q$ dependence of the form factors, and
particularly $A_2$, are not yet known with enough precision  to be conclusive.

%----------------------------------------
\subsection{Confronting the form factors to experimental ratios.}
\label{sec-confr}

We now turn to experiment in order to fix the remaining free parameters and to
know whether the data can be really understood in the above phenomenological
framework. We begin by the discussion of the ratios.

Although we would like to stick to our theoretical prejudices, eqs (\ref{sp})
and (\ref{one}), we still feel that the situation is very uncertain in the
subasymptotic regime. Therefore we shall also test other competing schemes in
order to know by comparison whether experiment is actually giving definite
indications on the \qq and mass dependence of the form factors. A few
definite lessons will be drawn but the exercise will not prove as conclusive
as we would have wished.

\subsubsection{Some simple prescriptions for mass and \qq dependence}
\label{sec-presc}

We intend to perform several $\chi^2$ fits of the ratios of form factors
according to the following method. We assume a given evolution prescription for
the dependence of the ratios of form factors as a function of the heavy mass at
$q^2_{max}$, and a given prescription for the behaviour of these ratios as a
function of $q^2$ for fixed masses. Next we combine in {\it one} $\chi^2$ fit
the experimental results for $D\ra K^{(\ast)} l \nu$ at $q^2=0$ and the
experimental results for the ratios $R$ and $R_L$ for $B\ra K^{(\ast)} \psi$.
The evolution prescriptions we have used are now described:

\begin{description}
\item {\bf 1) Soft-scaling-Pole} corresponds exactly to the assumption
(\ref{sp}), i.e. the heavy-to-heavy inspired scaling which implies a softened
scaling in heavy quark mass at fixed $\vec q$, and a ratio to $A_1(q^2)$ that
exhibits the kinematical pole similar to the
heavy-to-heavy scaling formulae.

\item{\bf 2) Soft-scaling-Constant} assumes the same softened scaling at
$q^2_{max}$ as before, but a ratio of form factors which stays constant as
$q^2$ varies for fixed masses:

\[{ \sqrt{4 m_{P_i} m_{P_f}} \over  m_{P_i} + m_{P_f} } \frac{ f_+(q^2)}{r_+} =
{ \sqrt{4 m_{P_i} m_{V_f}} \over  m_{P_i} + m_{V_f}  } \frac{V(q^2)}{r_V} =\]

\beq
= { \sqrt{4 m_{P_i}
m_{V_f}} \over m_{P_i} + m_{V_f} } \frac{A_2(q^2)}{r_2} = {\left ( m_{P_i} +
m_{V_f} \right )
\over \sqrt{4 m_{P_i} m_{V_f}} }{A_1(q^2) \over r_1 } = \eta(\vec q, m_f)
\label{sc}\eeq

\item{\bf 3) Hard-scaling-Pole} assumes hard scaling, i.e. the asymptotic laws
without $1/m_Q$ corrections. This is achieved by replacing in the ``soft-pole''
prescriptions at $q^2_{max}$:

 \[\sqrt{m_{P_i}}/(m_{P_i}+m_f) \to 1/\sqrt{m_{P_i}},\]
 but keeping the $q^2$ dependence at fixed masses as in (\ref{sp}). It
corresponds to:

\[ 2\left(\frac {  m_{P_f}}  { m_{P_i}} \right)^{\frac 1 2} { ( m_{P_i} +
m_{P_f} )^2  \over {4 m_{P_i} m_{P_f}}}\left[ 1 - {q^2 \over \left ( m_{P_i} +
m_{P_f} \right )^2}\right]
\frac{ f_+(q^2)}{r_+} =\]
\[
2\left(\frac {  m_{V_f}}  { m_{P_i}} \right)^{\frac 1 2}{ ( m_{P_i} + m_{V_f}
)^2  \over {4 m_{P_i} m_{V_f}}}\left[ 1 - {q^2 \over \left ( m_{P_i} +
m_{V_f} \right )^2}\right] \frac{V(q^2)}{r_V} =\]
\[
=2\left(\frac {  m_{V_f}}  { m_{P_i}} \right)^{\frac 1 2}{ ( m_{P_i} + m_{V_f}
)^2  \over {4 m_{P_i} m_{V_f}}}\left[ 1 - {q^2 \over \left ( m_{P_i} +
m_{V_f} \right )^2}\right] \frac{A_2(q^2)}{r_2} =\]
 \beq
\frac 1 2\left(\frac   { m_{P_i}} {  m_{V_f}}\right)^{\frac 1 2}{A_1(q^2) \over
r_1  } = \eta(\vec q, m_f)
\label{hp}\eeq

\item{\bf 4)  Hard-scaling-Constant} assumes the same hard scaling as above and
assumes that the ratios do not depend on $q^2$. It corresponds to:

\[ 2\left(\frac {  m_{P_f}}  { m_{P_i}} \right)^{\frac 1 2}
\frac{ f_+(q^2)}{r_+} =
2\left(\frac {  m_{V_f}}  { m_{P_i}} \right)^{\frac 1 2} \frac{V(q^2)}{r_V} =\]

\beq
= 2\left(\frac {  m_{V_f}}  { m_{P_i}} \right)^{\frac 1 2}\frac{A_2(q^2)}{r_2}
= \frac 1 2\left(\frac   { m_{P_i}} {  m_{V_f}}\right)^{\frac 1 2}{A_1(q^2)
\over r_1  } = \eta(\vec q, m_f)
\label{hc}\eeq

\end{description}

In the prescriptions 2) and 4), the assumption that the form factors have a
constant ratio in $q^2$ is inspired from the popular pole dominance
approximation\footnote{ In fact in the nearest pole dominance hypothesis, the
form factor ratios are not exactly constants since the position of the pole
depends on the form factor considered. But our aim here is  simply to exhibit
the main trends. Therefore we have used the
``constant ratio'' hypothesis for the sake of simplicity.}.

\subsubsection{Lessons from our global $\chi^2$ fits.}
\label{sec-less}

We now describe the results of our $\chi^2$ fits.
For the experimental results on $D\ra K^{(\ast)} l \nu$ at $q^2=0$ we take the
world average estimated by Witherell \cite{witherell}, and for $R_L$ we have
used in two differents fits the results from CLEO II \cite{cleo2} and from CDF
\cite{cdf}. For R we have used  $1.64\pm0.34$ from CLEO II \cite{browder}.

%___________________________________________________________________________
\begin{table}
\centering
\begin{tabular}{|c|c|c|c|c|c|} \hline
& & & & &  \\  extrapolation
&$\frac{\frac{A^{sb}_2(m_\psi^2)}{A^{sb}_1(m_\psi^2)}}{\frac{A^{sc}_2(0)}
{A^{sc}_1(0)}}$ & $\frac{V^{sc}(0)}{A^{sc}_1(0)}$ &
$\frac{A^{sc}_2(0)}{A^{sc}_1(0)}$ & $\frac{\Gamma_L}{\Gamma_{tot}}$ &
$\chi^2/dof$  \\
& & & & &  \\  \hline
Soft-Pole: CDF & 1.34 &  1.82 & 0.656 & 0.490 & 1.9  \\
Soft-Pole: CLEO & 1.34 & 1.67 & 0.520 &  0.567 & 9.0  \\
\hline
Soft-Cons: CDF & 1.77 &  1.75 & 0.556 &  0.386 & 5.7 \\
Soft-Cons: CLEO & 1.77  & 1.53 & 0.385 &  0.520 & 16.5 \\\hline
Hard-Pole: CDF & 2.14  & 1.70 &  0.480 &  0.322 & 9.5\\
Hard-Pole: CLEO & 2.14 & 1.43 &  0.292 &  0.498 & 22.5 \\ \hline
Hard-Cons: CDF & 2.83 &  1.66 & 0.375 &  0.233 & 16.1 \\
Hard-Cons: CLEO & 2.83  & 1.29 & 0.165 &  0.481 & 31.9 \\\hline
exp: AVE + CDF & - &  $1.9\pm .25$ & $0.74 \pm .15$&  $0.66\pm .14$  &-\\
exp: CLEO II & - &  - & -&  $0.80\pm .10$  &-\\
\hline
\end{tabular}
\caption{\it {The extrapolation procedures are explained in the text. In each
case, the values of the ratios of form factors $V^{sc}(0)/A_1^{sc}(0)$ and
$A_2^{sc}(0)/A_1^{sc}(0)$ have been
fitted to minimize the $\chi^2$ relative to the experimental numbers in the
last line. A two parameter fit for three constraints leaves one degree of
freedom (dof). The
experimental value for $R_L=\Gamma_L/\Gamma_{tot}$ is taken from CDF or CLEO
according to what is indicated in the first column. ``AVE'' refers to the world
average for the form factor ratios in \dk. Whenever it was  needed to combine
statistical and systematic errors, we have combined them in quadrature.}}
\label{tab-seul}
\end{table}
%-------------------------------------------------

Before discussing the outcomes of these fits, let us notice that in this
exercise, the result only depends on  the double ratios

 \[{\cal
R}_2=(A^{sb}_2(m_\psi^2)/A^{sb}_1(m_\psi^2))/(A^{sc}_2(0)/A^{sc}_1(0))\]
 \be {\cal
R}_+=(f^{sb}_+(m_\psi^2)/A^{sb}_1(m_\psi^2))/(f^{sc}_+(0)/A^{sc}_1(0))
\label{ratios}\ee
since within our Ansatz, the double ratio ${\cal R}_V$ satisfies
\[{\cal
R}_V=(V^{sb}(m_\psi^2)/A^{sb}_1(m_\psi^2))/(V^{sc}(0)/A^{sc}_1(0))={\cal
R}_2.\]

The values of these double ratios depend both on the mass dependence of the
form factor ratios, i.e. the corrections to asymptotic scaling, and on the
$q^2$ dependence of these ratios.

\paragraph{ We have first performed a fit restricted to the $K^\ast$ final
state,} leaving aside the $f_+$ form factors and the ratio \gs. The results are
displayed in table \ref{tab-seul}.  As a first conclusion from table
\ref{tab-seul} one sees that the best fit is Soft-scaling-Pole. The reason for
that is that
the large experimental values of \glgt and $A^{sc}_2(0)/A^{sc}_1(0)$ impose the
double ratio ${\cal R}_2$ to be rather small, as argued in section
\ref{sec-failu}, thus suggesting either that the asymptotic scaling law
($A_2/A_1 \propto m_Q$) is strongly softened, or that the $A_2/A_1$ ratio
decreases dramatically with decreasing $q^2$, or some compromise between both
effects.
Indeed the Soft-scaling-Pole Ansatz has both a softened scaling at \qmax and a
decrease of the $A_2/A_1$ ratio  with decreasing $q^2$. Therefore, this Ansatz
yields the smallest ratio ${\cal R}_2$, and thus a larger \glgt .
A  roughly acceptable fit is thus obtained for CDF data, with a confidence
level of $\sim 20\%$, but not for CLEO. To obtain an acceptable fit with CLEO
data, one would need a value of ${\cal R}_2$ sensibly smaller than 0.8. Such a
low value seems very difficult to obtain in any natural way. Table
\ref{tab-seul} also shows that Soft-scaling is generally favored and
Hard-scaling is in very strong disagreement with data.

\paragraph{In a second step we have added to our fits the data concerning the
$K$ final state.} The results are displayed in table \ref{tab-cdf}.
Qualitatively there is no big change. Looking in more detail, it appears that
the best fits now include on the same level the Soft-scaling-Pole and the
Soft-scaling-Constant cases, with a worst confidence level (around 2\%).
The reason for that is that all our Ans\"atze correspond to ${\cal R}_+\simeq
{\cal R}_2$, and, while the data on \glgt require a small ${\cal R}_2$ as just
argued, the data on \gs and $f^{sc}_+(0)/A^{sc}_1(0)$  on the contrary require
a not too small double ratio ${\cal R}_+$. This can be understood as follows.

%___________________________________________________________________________
\begin{table}
\centering
\begin{tabular}{|c|c|c|c|c|c|c|c|c|} \hline
& & & & & &  &  & \\  extrapolation
&$\frac{\frac{A^{sb}_2(m_\psi^2)}{A^{sb}_1(m_\psi^2)}}
{\frac{A^{sc}_2(0)}{A^{sc}_1(0)}}$ &
$\frac{\frac{f^{sb}_+(m_\psi^2)}{A^{sb}_1(m_\psi^2)}}{\frac{f^{sc}_+(0)}{A^{sc}_1(0)}}$
&  $\frac{f^{sc}_+(0)}{A^{sc}_1(0)}$& $\frac{V^{sc}(0)}{A^{sc}_1(0)}$  &
$\frac{A^{sc}_2(0)}{A^{sc}_1(0)}$  & $\frac{\Gamma(K^\ast)}{\Gamma(K)}$ &
$\frac{\Gamma_L}{\Gamma_{tot}}$ &
$\chi^2/dof$  \\
& & & & & &  &  & \\  \hline
Soft-Pole: CDF & 1.34 & 1.28 & 1.45 & 1.62 & 0.809 & 2.15 & 0.449 & 4.2  \\
Soft-Pole: CLEO & 1.34 & 1.28 & 1.47 & 1.45 & 0.680 & 2.21 & 0.533 & 8.6  \\
\hline
Soft-Cons: CDF & 1.77 & 1.70 & 1.32 & 1.66 & 0.600 & 1.81 & 0.375 & 3.2 \\
Soft-Cons: CLEO & 1.77 & 1.70 & 1.34 & 1.43 & 0.443 & 1.87 & 0.510 & 8.9
\\\hline
Hard-Pole: CDF & 2.14 & 2.14 & 1.25 & 1.72 &  0.472 & 1.60 & 0.323 &  4.7 \\
Hard-Pole: CLEO & 2.14 & 2.14 & 1.26 & 1.43 &  0.293 & 1.64 & 0.498 & 11.2\\
\hline
Hard-Cons: CDF & 2.83 & 2.83 & 1.20 & 1.79 & 0.344 & 1.44 & 0.234 & 8.4 \\
Hard-Cons: CLEO & 2.83 & 2.83 & 1.20 & 1.40 & 0.123 & 1.44 & 0.481 & 16.4
\\\hline
exp: AVE + CDF & - & - & 1.26 & 1.9 & 0.74 & - & 0.66 &-\\
  & - & - & $\pm .12$ & $\pm .25$ & $\pm .15$ & -& $\pm .14$ &-\\
exp: CLEO & - & - & -& - & - & 1.64 & 0.80 &-\\
  & - & - & -& - & - & $\pm .34$& $\pm .10$ &-\\
\hline
\end{tabular}
\caption{\it {The extrapolation procedures are explained in the text. In each
case, the values of the ratios of form factors $V^{sc}(0)/A_1^{sc}(0)$,
$A_2^{sc}(0)/A_1^{sc}(0)$ and $f_+^{sc}(0)/A_1^{sc}(0)$ have been
fitted to minimize the $\chi^2$ relative to the experimental numbers in the
last line. A three parameter fit for five constraints leaves two degrees of
freedom (dof). The
experimental value for $R_L=\Gamma_L/\Gamma_{tot}$ is taken from CDF or CLEO
according to what is indicated in the first column. ``AVE'' refers to the world
averages for form factor ratios in \dk. Whenever it was needed to combine
statistical
and systematic errors, we have combined them in quadrature.}}
\label{tab-cdf}
\end{table}
%-------------------------------------------------

 The ratio $R$ (\ref{rstar}) is given by:

\be R=1.081\left(\frac { A_1^{sb}(m_\psi^2)}{f_+^{sb}(m_\psi^2)}\right)^2
\left
\{{2\left[1+0.189\left(\frac{V^{sb}(m_\psi^2)}{A_1^{sb}(m_\psi^2)}\right)^2
\right] +\left(3.162 -1.306 \frac {A_2^{sb}(m_\psi^2)}{A_1^{sb}(m_\psi^2)}
\right)^2}\right\}\label{r}\ee

Multiplying the l.h.s. of eqs (\ref{rlform}) and (\ref{r}) we get:

\be R(1- R_L) = 2.162\left(\frac
{A_1^{sb}(m_\psi^2)}{f_+^{sb}(m_\psi^2)}\right)^2
\left[1+0.189\left(\frac{V^{sb}(m_\psi^2)}{A_1^{sb}(m_\psi^2)}\right)^2\right]
\label{beau}\ee

This gives obviously a lower bound on $f_+/A_1$. For the conservative upper
bounds of $R\le 2.5$ and $1-R_L\le 0.5$  and setting still more conservatively
$V$ to zero we get $f_+/A_1 \ge 1.32$. For a more realistic estimate,
$V/A_1\simeq 2$, and $R\le 2.0$ we get ${\cal R}_+ \ge 1.60$.
Contrarily to our discussion in section \ref{sec-failu},  we find here a lower
bound which in itself is compatible with the hard scaling behaviour but not
with such a soft scaling as required for $A_2/A_1$ (remember we had $A_2/A_1\le
1.3$  for $V=0$ and for a more realistic $V/A_1$, $A_2/A_1\le 1$). {\it Clearly
the trend for $f_+/A_1$ is somewhat opposite to the one for $A_2/A_1$ }.

The $\chi^2$ fits tried a compromise between these two opposite trends, and
this is why the Soft-scaling-Constant Ansatz now fits as well as the
Soft-scaling-Pole one: although the prediction for \glgt is worst for the
former, it gives a better \gs ratio, since it corresponds
to larger double ratios.
Again the Hard-scaling cases are rejected.
The two best fits are hardly acceptable in the case of the CDF value $R_L=
0.66\pm 0.14$ and fail with CLEO II's much more restrictive value
$R_L=0.80\pm 0.095$ (as it would have failed with the Argus bound $R_L>0.78,
95\%$ confidence level).
There is a real difficulty, as noted in \cite{gourdin}, to account for the data
on \dk and \bpsi.

\paragraph{What could be the way out of this dilemma ?} One may of course
question the factorization hypothesis which is the basic hypothesis in all this
paper. Although we are thoroughly convinced that the factorization hypothesis
may very reasonably be doubted in a $1/N_c$ subdominant channel as is $B\ra
K^{(\ast)}\psi$, we decided to leave all this discussion outside the present
paper. We may also hope that more precise experiments will evolve in a
direction that will make the problem not so acute. Although the $\chi^2$
 may seem horrific when
the CLEO II value for $R_L$ is used, it should be kept in mind that
a small variation of the experimental value may lead to a dramatic decrease of
the $\chi^2$. A comparison with CDF gives a first example of that.

Finally our hypothesis, displayed in eqs (\ref{sp}), (\ref{sc})-(\ref{hc}),
leading
$V(q^2)/A_1(q^2)\propto A_2(q^2)/A_1(q^2)$ and $f_+(q^2)/A_1(q^2)\propto
A_2(q^2)/A_1(q^2)$\footnote{The latter equation is only approximately valid in
eqs. (\ref{sp}) and (\ref{sc})  due to the $K-K^\ast$ mass difference.} may
also be criticised. It would of course be meaningless to relax these
constraints in the above-described $\chi^2$ fits, since we would have too many
free parameters. Qualitatively it is obvious that any prescription with ${\cal
R}_+$ sensibly larger than ${\cal R}_2$ would lessen the $\chi^2$. But we see
no sign of such a trend in the models we have considered. Another lessening of
the $\chi^2$ would happen if ${\cal
R}_V=(V^{sb}(m_\psi^2)/A^{sb}_1(m_\psi^2))/(V^{sc}(0)/A^{sc}_1(0))$ was
sensibly smaller than ${\cal R}_2$. This happens to be the case, although in a
quantitatively insufficient amount, in the Orsay Quark Model (see section
\ref{sec-oqm}): in eq. (\ref{v}) it appears that $V(q^2)$ contains a relatively
large corrective factor $Y$, (\ref{xy}), which decreases with the initial mass.
This term tends to decrease all the $\chi^2$ in table \ref{tab-cdf} but not
enough. It has the effect of providing for the form factor $V(q^2)$ an even
larger softening of the increase predicted by the asymptotic scaling law
(\ref{hl}). It is interesting to notice that such a large correction to the
asymptotic scaling law for $V(q^2)$ in the direction of a softening has been
found in lattice calculations \cite{abada}, although the large statistical
errors in these calculations do not allow to draw yet a final conclusion.
Finally,  our difficulties to get small $\chi^2$ is not too surprising since
the $\chi^2$ tends anyhow to become large when experimental errors decrease,
unless a very accurate model is available, which is certainly not the case in
the present attempts.

\paragraph{To summarize,} we have found, using our simple phenomenological
Ans\"atze for the dependence of the form factors ratios in \qq and in masses,
that:
\begin{itemize}

\item The experiment comparison between \bpsi and \dk favors a soft \hl
scaling, eqs. (\ref{sp}) and (\ref{sc}).

\item The best \qq dependence cannot be selected from this analysis alone,
although the separated phenomenological study of the $K^\ast$ final states
(table \ref{tab-seul}), as well as several theoretical considerations, tend to
favor the existence of the ``kinematical pole'' as in eq. (\ref{sp}).

\item There remains a difficulty to reconcile experimental results in \bpsi
and \dk when taking CDF results for $R_L$ ($\chi^2/dof \simeq 3$) which grows
even worst when using CLEO or ARGUS values for $R_L$. There seems to be also a
particular difficulty to fit simultaneously $R$ and $R_L$. Only fragile
indications of possible ways out these difficulties are known today.

\end{itemize}

%-------------------------------------------
\subsection{\qq  dependence of $A_1(q^2), f_+(q^2)$, etc. from experiment.}
\label{sec-abs}

Up to now we have mainly considered the ratios of form factors,
$A_2/A_1, V/A_1$ and $f_+/A_1$. In this subsection we try to go beyond and
consider how  the form factors themselves depend on \qq. We shall now gather
from
different sources information about $A_1(q^2)$, $f_+(q^2)$, and we shall see
that these combined informations are rather compatible with what we already
know about the ratios.

Although pure phenomenology using combined final $K$ and $K^\ast$ data did not
allow us to choose between Soft-scaling-Pole (\ref{sp}) and
Soft-scaling-Constant (\ref{sc}) Ans\"atze,  the separated phenomenology
for the $K^\ast$ final state and also several theoretical arguments lead us to
chose
the Soft-scaling-Pole.

Of course the above rough agreement of Soft-scaling-Pole for \glgt and for \gs
does not depend of the  value of $\eta(\vec q,m_f)$ in eq. (\ref{sp}). This
subsection is devoted to argue in favor of our choice in eq. (\ref{one}) for
$\eta(\vec q,m_f)$.
The meaning of eq. (\ref{one}) is in fact that we choose $A_1(q^2)$ to be a
constant:

 \be A_1(q^2)=r_1 \,\frac{\sqrt{ 4 m_{P_i} m_{V_f}}}{
m_{P_i}+m_{V_f}}.\label{a1con}\ee
Of course, only the product $r_1\, \eta(\vec q,m_f)$ is relevant, not the
separate values of $r_1$ and $\eta(\vec q,m_f)$. Next, let us stress that our
QMI Ansatz, eqs (\ref{sp}) and (\ref{one}), does not mean that we believe
$A_1(q^2)$ to be a constant. We are indeed sure, from its analytic properties
(the axial current singularities in the $t$-channel),
that  $A_1(q^2)$ is {\it not} a constant.
Our Ansatz really  means that we believe $A_1(q^2)$ {\it to vary slowly with
\qq} in the physically relevant region, at least slower than predicted from
pole dominance. Eq. (\ref{one}) is only the simplest possible Ansatz to express
this feature of a slow variation. Let us now summarize a few arguments in favor
of the slow variation of $A_1(q^2)$.

\paragraph {i) Pole-like behavior of $D \to Kl\nu$.}
In \cite{witherell} it is argued that $f_+(q^2)$ in \dk decay may well be
fitted by a vector meson pole, and the fitted pole mass is $M^\ast= (2.00 \pm
0.11 \pm 0.16)$ GeV, in good agreement with the value of $2.1$ GeV expected for
the mas of the $D^\ast_s$ meson. This fit does not establish the detailed
analytic form of $f_+(q^2)$ since an exponential fit is told to agree as well.
But it certainly conveys the message of a $f_+(q^2)$ increasing with \qq like a
pole term rather than, say, a dipole or a constant. Combined with both eq.
(\ref{sp}) this \qq dependence points toward a constant $\eta(q^2,m_K)$
\footnote{One may object that the pole of $f_+(q^2)$ in our QMI model is at
$m_D+m_K \simeq 2.3$ GeV, sensibly larger than the fitted pole
above-mentioned. This  objection is obvioulsy valid, it is obvious that eq.
(\ref{sp}) and (\ref{one}) do not provide the correct analytic properties to
the form factors, they do not have the wanted poles and cuts, etc. For sure,
our Ans\"atze are only approximations that we hope to be good enough in the
physical region.}.

\paragraph{ ii) The phenomenological factorization coefficient $a_2$.} Although
there is no theoretical principle to fix $a_2$ in the phenomenological BSW
factorization prescription, it seems reasonable that it cannot be too different
from its
value, $a_2^{SVZ}\simeq 0.1$, in the standard SVZ factorization, eq.
(\ref{a1a2}). The
results for our favored Soft-scaling-Pole are displayed in table
\ref{tab-qstar}. It appears that the double pole assumption gives very large
values for $a_2$, while the constant behaviour for $A_1$ is favored as it
gives the smallest $a_2$ (remember that this corresponds to a pole behaviour of
$f_+$). This confirms our choice (\ref{one}).
Still our preferred fitted $a_2$, ranging from 0.22 to 0.28, might be
considered as rather large compared to the SVZ value.

%___________________________________________________________________________
\begin{table}
\centering
\begin{tabular}{|c|c|c|c|c|} \hline
&  &  & &\\
  Model &   $\frac{\Gamma(K^\ast)}{\Gamma(K)}$ &
$\frac{\Gamma_L}{\Gamma_{tot}}$ &  $a_2$ for $K$ & $a_2$ for $K^\ast$ \\
& & & & \\  \hline
BSW I \cite{bsw1} & 4.23 & 0.57 & 0.39 & 0.24 \\ \hline
BSW II \cite{bsw2} & 1.61 & 0.36 & 0.26 & 0.26  \\ \hline
Soft-Pole: $A_1$-Pole & 4.14 & 0.45 & 0.37-0.43 & 0.30-0.35 \\ \hline
Soft-Pole: $A_1$-Const. & 3.21 & 0.45 & 0.25-0.28 & 0.22-0.25 \\ \hline
\end{tabular}
\caption{\it The fitted values of the phenomenological factorization parameter
$a_2$ from $B\to \psi K$ branching ratios are given in column four, the one
fitted from $B\to \psi K^\ast$ in the last column. The first two lines use BSW
models. The starting point for the other lines are the Soft-Pole form factor
ratios in table 3, both with CLEO and CDF values for $R_L$.
Given the ratios, either we take $A_1$ or $f_+$ from \dk experiment. In the
last two columns we report the range of fitted $a_2$ obtained with four
different choices: $A_1$ or $f_+$ from experiment, CLEO or CDF for $R_L$. It
appears that these ranges are narrow enough.
An additional prescription is used for the $A_1$ dependence on \qq: pole
dominance
or constant. The corresponding indication is given in a transparent way in the
first column.}
\label{tab-qstar}
\end{table}
%-------------------------------------------------

\paragraph{ iii) Orsay Quark Model.}
OQM will be detailed in section \ref{sec-oqm}. It seems to suggest a weak $q^2$
dependence for $A_1(q^2)$, related to the slow variation of a Lorentz
contracted overlap near $q^2=0$, and an approximate pole behavior for the rest
of the form factors.

\paragraph{ iv) Lattice calculations and QCD Sum Rules.} Lattice and QCD sum
rule results will be commented in sections \ref{sec-latt} and \ref{sec-qcd}.
The situation is not so clear, errors are still large in lattice calculations,
different QCD Sum Rules calculations show discrepancies. However there is a
converging set of indications against a pole dominance in the case of $A_1$,
generally pointing towards a flatter \qq dependence.

\subsection{Mass dependence of form factors at $q^2=0$.}

We have already noticed, eq. (\ref{zerohl}), that our Ansatz (\ref{sp}) for the
form factor ratios implies a very simple mass dependence at \qqz. In this
section we will draw the consequences of  our different Ans\"atze  on
the mass dependence of the form factors at \qqz.

\paragraph{Hard-scaling-Constant with a pole for $A_1$.}

This prescription, using eq (\ref{hc}) and
\be
	A_1(q^2)={A_1(0) \over 1 -
{q^2 \over m_{B_s^{\ast\ast}}^2}},\label{a1pole}
\eeq
is equivalent to assuming a pole dominance for all form factors, as was done in
\cite{bsw1} and \cite{altomari}. At \qqz one obtains:

	\[{f^{sb}_+(0) \over f^{sc}_+(0)},\quad {V^{sb}(0) \over V^{sc}(0)},\quad
{A^{sb}_2(0) \over A^{sc}_2(0)} = \left ( {m_{D} \over m_B} \right )^{{1 \over
2}}\left(1+O\left(\frac \Lambda{m_D}\right)\right)\]

\beq
{A^{sb}_1(0) \over A^{sc}_1(0)} = \left ( {m_{D} \over m_B} \right )^{{3 \over
2}}\left(1+O\left(\frac \Lambda{m_D}\right)\right).\label{hcpz}
\eeq

\paragraph{Hard-scaling-Pole with a pole for $A_1$.}

This prescription starts from (\ref{hp}) also with (\ref{a1pole}). It yields
double poles for the other form factors than $A_1$. It was used in \cite{bsw2}.
At \qqz it gives the same ratio for all form factors:

\beq
{f^{sb}_+(0) \over f^{sc}_+(0)},\quad {V^{sb}(0) \over V^{sc}(0)},\quad
{A^{sb}_1(0) \over A^{sc}_1(0)},\quad
{A^{sb}_2(0) \over A^{sc}_2(0)} = \left ( {m_D \over m_B} \right )^{{3 \over
2}}\left(1+O\left(\frac \Lambda{m_D}\right)\right).\label{hppz}
\eeq

\paragraph{Hard-scaling-Pole with a constant for $A_1$.}

Eq. (\ref{hp}) with (\ref{one}). It gives at \qqz also the same ratio for all
form factors, with a different power:

	\beq
{f^{sb}_+(0) \over f^{sc}_+(0)},\quad {V^{sb}(0) \over V^{sc}(0)},\quad
 {A^{sb}_1(0) \over A^{sc}_1(0)},\quad
{A^{sb}_2(0) \over A^{sc}_2(0)} = \left ( {m_D \over m_B} \right )^{{1 \over
2}}\left(1+O\left(\frac \Lambda{m_D}\right)\right).\label{hpcz}
\eeq

The three preceding cases have only an asymptotic validity, for $m_B, m_D
\to \infty$. In eqs (\ref{hp}) and (\ref{hc}), the corrections have been
retained, which explains an $O(1/m_D)$ difference between the latter and eqs.
(\ref{hcpz})-(\ref{hpcz}). On the contrary, the next equation retains the
non-asymptotic corrections.

\paragraph{Soft-scaling-Pole with a constant for $A_1$.}

Eq. (\ref{sp}) with (\ref{one}). As we have already told, this is our prefered
Ansatz. It gives at \qqz

\bea
{f^{sb}_+(0) \over f^{sc}_+(0)} & = & \left ( {m_B \over m_D} \right )^{{1
\over 2}} \left ( {m_D +
m_K \over m_B + m_K} \right ) \nonumber \\
{V^{sb}(0) \over V^{sc}(0)} & = & {A^{sb}_1(0) \over A^{sc}_1(0)} =
{A^{sb}_2(0) \over A^{sc}_2(0)} =
\left ( {m_B \over m_D} \right )^{{1 \over 2}} \left ( {m_D + m_{K^{\ast}}
\over m_B + m_{K^{\ast}}}
\right ).\label{spcz}
\eea
which reduce of course to (\ref{hpcz}) in the asymptotic regime.

The results (\ref{hcpz}) to (\ref{spcz}) will be useful in section
\ref{sec-disc} to discuss the models.

%*********************************************
\section{Discussion of theoretical approaches.}

\label{sec-disc}

The discussion in section \ref{sec-class} provides us with some tools to look
further into the theoretical schemes con\-si\-de\-red in the
literature:

\begin{enumerate}
\item Lattice calculations (LC).
\item QCD sum rules (QCDSR).
\item Pole model of Bauer, Stech and Wirbel \cite{bsw1}.
\item Pole-dipole model of Neubert et al. (NRSX)
\cite{bsw2}.
\item VMD model of Altomari and Wolfenstein \cite{altomari}.
\item Quark model of Isgur, Scora, Grinstein and Wise (ISGW) \cite{isgw}.
\item Orsay quark model (OQM) \cite{oqm}.

\end{enumerate}

In examining all these approaches, we will pay attention to two main aspects :

i) To what extent are they fulfilling the asymptotic theorems, including the
\hhs when both masses are heavy ?

ii) Why do they fail at explaining the  $B \to \psi K(K^{\ast})$   data ?

\subsection{Lattice complemented with $q^2$ Ansatz.}
\label{sec-latt}

We have used the lattice Monte Carlo calculations of the form factors performed
by the European Lattice Collaboration at $\beta=6.4$. The details on the
lattice parameters can be found in ref. \cite{abada}.  What is relevant here is
that the lattice spacing is large enough to allow relatively large quark
masses, from which to extrapolate up to the $B$ meson. Reversely, the
statistics is not too high, leading to large errors. For the light quark we
have used the  value $\kappa=0.1495$ which happens to be very close to the
``physical'' strange quark: $\kappa_s=0.1495\pm 0.0001$ \cite{abada}. The
description of the extrapolation in the heavy quark mass up to the $b$ quark is
to be found in \cite{abada}.  But we have modified the extrapolation
procedure in \qq. In \cite{abada} a pole dominance approximation was used for
all form factors. Since we have strong reasons exposed in this paper to doubt
this hypothesis, we have done the following:

The lattice calculations have been performed for the $A_1$ form factors at five
different values of \qq. However, due to the statistical noise, we have only
used the three closest to \qmax. We then perform a two parameter fit for $A_1$:

\be A_1(q^2)= a + \frac {b}{q^2-M_p^2}\label{pc}\ee
or equivalently

\be A_1(q^2)= \frac {a q^2 + c}{q^2-M_p^2},\quad c=b-a M_p^2\ee
where $M_p$ is the lattice mass of the lightest t-channel axial meson pole.
The justification of such a form is twofold: i) the form factor must indeed
present a pole at $q^2=M_p^2$; ii) the constant $a$ mimics the subtraction
constant of the dispersion relation.

To present the results of our \qq dependence fit of $A_1(q^2)$ we will define a
ratio

\be P_1=\left(\frac {M_p^2-q^2_{max}}{A_1(q^2_{max})}\right) \left.
\frac{\partial A_1(q^2)}{\partial q^2}\right \vert_{q^2_{max}}\label{p1}\ee
such
that $P_1=1$ in the pole dominance hypothesis. Using two different analysis
methods explained  in \cite{abada} we find $P_1=0.38 \pm 0.50$ ($0.92 \pm
0.41$) for the ``analytic'' (``ratio'') method. One may see an indication in
the direction of a flatter behaviour of $A_1$ than predicted by the pole
dominance, but the errors prevent any firm statement.

Concerning the other form factors, $A_2, f_+, V$, the  lattice calculations do
not give a direct estimate at \qmax, and the same fitting procedure is not
possible. We then have chosen to assume that the \qq dependence of the ratios
$A_2(q^2)/A_1(q^2)$, $f_+(q^2)/A_1(q^2)$ and $V(q^2)/A_1(q^2)$ are given by eq.
(\ref{sp}) or by eq. (\ref{sc})\footnote{Let us repeat that only the $q^2$
dependence is taken from eqs. (\ref{sp}) and (\ref{sc}), the mass dependence
has been fitted from the lattice as explained in \cite{abada}.}

Our results are reported in table \ref{tab-lat}. Clearly the errors are
overwhelming for the ratio $R$, but the results for the ratio $R_L$ delivers a
clear message in favor of the ``Soft-scaling-Pole'' prescription for the \qq
dependence of form factors ratios. With the latter prescription, the lattice
results is within $1 \sigma$ from CDF, but $3 \sigma$ from CLEO II.

%___________________________________________________________________________
\begin{table}
\centering
\begin{tabular}{|c|c|c|} \hline
& &  \\ form factors ratios   & $\frac{\Gamma(K^\ast)}{\Gamma(K)}$ &
$\frac{\Gamma_L}{\Gamma_{tot}}$   \\
& & \\  \hline
Soft-Pole eq.(\ref{sp})& $ 3.5\pm 2.5$ & $0.47\pm 0.11 $\\ \hline
Soft-Cons. eq.(\ref{sc})& $ 1.9\pm 1.4$ & $0.27\pm0.16$ \\ \hline
exp. &$ 1.64\pm 0.34$ &$ 0.64\pm 0.14 \,(0.80\pm 0.10)$ \\
\hline
\end{tabular}
\caption{\it{ The results from lattice calculations at $\beta=6.4$. The mass
dependence of the form factors and the $q^2$ dependence of $A_1(q^2)$ has been
fitted as explained in the text. The $q^2$ dependence of the ratios $A_2/A_1,
f_+/A_1$ and $V/A_1$ have been taken according to a prescription indicated in
the
first column. The experimental number  have been taken from CDF, those from
CLEO II  being indicated in brackets. }}
\label{tab-lat}
\end{table}
%---------------------------------------------------------------------

Preliminary results from APE lattice collaboration \cite{ape} on $B\to K^\ast
\gamma$
seem to indicate also an increase of $T_2(q^2)$ with \qq much slower than
expected from pole dominance, i.e. the analogous of parameter $P_1$ defined in
eq (\ref{p1}) looks much smaller than 1. This is interesting in view of the
fact that $T_2$ is asymptotically equal to $A_1$. Similarly, $T_1$ is
asymptotically equal to $V$.
Now, APE collaboration finds a much faster increase of $T_1$ with \qq than
$T_2$, as would be predicted from an extension of eq (\ref{hhbsg}) to the \hl
system.

\subsection{QCD sum rules.}
\label{sec-qcd}

There has been several studies \cite{dosch}-\cite{narison} of the \qq
dependence of \hl form factors using different varieties of QCD sum rules:
Laplace sum rules, hybrid sum rules and light-cone sum rules. Some kind of
consensus seems to have emerged, that we could characterize by saying that
these authors find an agreement with vector meson dominance for vector current
form factors, and a more gentle slope for $A_1$. Still, when one looks in more
detail, the different predictions for $A_1$ differ somehow. Ali et al.
\cite{ali} find a
softly increasing  $A_1$ for all \qq, while Ball \cite{ball} finds $A_1$
decreasing
with \qq for $q^2\le 15. \mbox{GeV}^2$. Narison \cite{narison} finds a
decreasing
$A_1$ in the $q^2\le 0$ region, and catches up with Ball's result. Ball finds
an increasing $A_2$ in the same region $q^2\le 15. \mbox{GeV}^2$ where $A_1$
decreases, and interestingly enough, at a first glance the plots show that the
in
this \qq region, the ratios $A_2(q^2)/A_1(q^2)$ and $V(q^2)/A_1(q^2)$ are not
very different from the ``kinematical pole'' term $ 1/(1 - q^2 /( m_{P_i} +
m_{V_f} ))^2$ (see eq. (\ref{sp})). Unhappily, for the limited $q^2\ge 15\,
\mbox{GeV}^2$ region,
the trend is reversed and the ratio $A_2(q^2)/A_1(q^2)$ even starts decreasing.
Ali et al. also find a $V(q^2)/A_1(q^2)$ ratio that has some analogy with the
``kinematical pole'' in the whole region they plot: $q^2\le 17 \,\mbox{GeV}^2$.
Finally, comparing Belyaev at al \cite{belyaev} to Ali et al. \cite{ali} we see
that
$f_+(q^2)/A_1(q^2)$ also increases with \qq, although maybe in a milder way
than
$V(q^2)/A_1(q^2)$. Finally, let us mention that Narison \cite{narison} finds
asymptotically, when the mass really goes to infinity ($m_Q\gg m_b$), an
analytic evidence for a pole behaviour of the form factor ratios  $A_2/A_1,
V/A_1$ and $f_+/A_1$, but this happens through a polynomial decrease of $A_1$
and a constancy of the other form factors.

To summarize, notwithstanding sensibly different predictions for $A_1$, there
is an almost general agreement (except for a small  domain near \qmax in
\cite{ball})
on an increase of the form factor ratios, rather similar to the ``kinematical
pole'' behaviour in (\ref{sp}).

Using the results obtained by Patricia Ball, \cite{ball} and \cite{ballp} for
$B\to \pi, \rho$, and assuming they are also valid for $B\to K^{(\ast)}$, we
have computed the ratios $R$ and $R_L$ that are reported in table
\ref{tab-RRL}. $R_L$ comes out rather small due to a too large value:
$A_2(m_\psi^2)/A_1(m^2_\psi)\simeq 1.2$: indeed, given the value
$V(m_\psi^2)/A_1(m^2_\psi)\simeq 2.8$ in \cite{ballp}, the constraint $R_L\ge
0.5$ translates into $A_2(m_\psi^2)/A_1(m^2_\psi)\le 0.8$. The ratio $R$ comes
out too large due to a too small value $f_+(m^2_\psi)/A_1(m^2_\psi)\simeq 1$,
while a reasonable lower bound of $\sim 2$ may be derived from eq.
(\ref{beau}).  We have neglected $SU(3)$ breaking
which could change the results, but we doubt this change could be large enough
to recover an agreement with \bpsi data. Once more we see how difficult a
challenge these data are for all known theoretical approaches.

\subsection{Quark Models.}
\label{sec-quarks}

Under this heading are included very different approaches. This wide range of
methods reflects primarily the unability of the simple-minded non relativistic
model to describe the form factors: this unability consists in two main facts:
i) Away from \qmax the model is highly ambiguous, as one soon reaches
relativistic velocities. ii)  The slope $\rho^2$ of the Isgur-Wise function is
definitely too small. Various attempts have been made to cure these defects,
either by appealing to ideas connected with vector meson dominance (VMD) or
with the relativistic effects \cite{gif-91}, \cite{marbella-93}. We have
restricted ourselves
to the more extensively used models in literature, leaving aside very
interesting ones such as the work by Jaus and Wyler \cite{wyler}.

As a general remark, it must be emphasized that all quark models except the
OQM,  section \ref{sec-oqm}, do not satisfy the \hhs when both $m_{P_i}$ and
$m_f$ are large,  and in some  cases (BSW models) violate also \hls, as we
shall argue in subsection \ref{sec-bsw}.
Also, on the empirical side, all models fail to explain $D \to K^\ast$ decays.
The axial form factors are too large, resulting in a too large $\Gamma(D\to
K^\ast l\nu)/\Gamma(D\to K l\nu)$ ratio. This is easily understandable by the
fact that no attempt has been made to incorporate in them the binding effects
which are crucial in obtaining a relative reduction of axial with respect to
vector  form factors (remember $D\to K$ is a purely vector transition).  A
similar situation was discussed in the past about the nucleon axial vector
coupling, the $G_A/G_V$ (sometimes noted $g_1/f_1$) ratio \cite{lopr}.

\subsubsection{BSW models.}
\label{sec-bsw}

The BSW models  have been used extensively to analyze non-leptonic $D$
and $B$ decays with the help of the additional BSW factorization assumption.
There are two main and very different ingredients in these models, which should
not be confused :

i) A standard quark model, which is used only at $q^2=0$.
The values at $q^2 = 0$ are found to be approximately the same for all $c \to
s$ and
for all $b \to s$ form factors:

\beq
f^{sc}_+(0) \simeq V^{sc}(0) \simeq  A^{sc}_1(0) \simeq
A^{sc}_2(0) \simeq 0.8
\label{d0}\eeq

\beq f^{sb}_+(0) \simeq V^{sb}(0) \simeq A^{sb}_1(0) \simeq A^{sb}_2(0) \simeq
0.35.\label{b0}
\eeq

One notes that the values of the  form factors $V$ and $A_2$ in (\ref{d0}) are
not
consistent with what is now known from experimental $D$ semi-leptonic decays
(\ref{dexp}) and (\ref{drexp}).

{}From eq. (35) in \cite{nr}, one can deduce the common asymptotic behavior of
all the form factors at $q^2=0$ as function of $m_{P_i}$:

\be h \propto \frac 1{m_{P_i}}\label{nr}\ee

ii) Different possible Ans\"atze about the \qq dependence away from 0, which
lead to two distinct phenomenological models: either a pole for each form
factor (BSW I), \cite{bsw1}, or a dipole for some and a pole for others (BSW
II), \cite{bsw2}. These Ans\"atze are probably motivated by the feeling that
na\"{\i}ve application of the quark model would fail, and also by the general
idea of pole dominance; the latter reason is why one may speak of ``hybrid''
models. It must be said that in the case of BSW II, this VMD idea is in
addition mixed with still another idea perhaps contradictory to it, the
squaring of the pole, which is inspired from Isgur-Wise \hhs.

In both cases the asymptotic behavior at $q^2=0$ as function of the heavy mass
will be found below to be contradictory with what could be deduced from \hls
and the assumed \qq dependence. This implies that the models {\it do not
fulfill the \hls properties}.

\subsubsection{First BSW model.}

 The BSW single pole model \cite{bsw1} for all the form factors has been used
to analyze the non-leptonic $D$ and $B$ decays.
 Note that, independently of the precise value of the ratio between the
numbers, the ratio between B and D form factors is roughly identical for all
form factors. This seems hardly compatible with the relations (\ref{hcpz}),
which
result from the combination of hard scaling and the \qq dependence assumed in
the
BSW I Model, and
would imply  asymptotically a different B to D ratio for $f_+$, $V$, $A_2$, and
$A_1$ respectively. This suggests that the model
does not fulfill the \hl scaling properties. One can indeed prove  rigorously
this fact in the asymptotic limit $m_{P_i}\gg m_f$, by observing that the
asymptotic behavior (\ref{nr}) is in contradiction with the asymptotic
relations (\ref{hcpz}) deduced from \hl scaling.

The BSW I model gives $R_L = 0.59$; this is not too bad, but the ratio $R =
4.23$ is much too large (see table \ref{tab-RRL}). This is due to the fact that
the ratio $f_+/A_1$ is too small as seen in the table \ref{tab-RRL}.

\subsubsection{Second BSW or NRSX model.}

The pole-dipole model of Neubert, Rieckert, Stech and Xu (NRSX) \cite{bsw2}
uses, to our
knowledge, the same values (\ref{d0}) and (\ref{b0}) at $q^2 = 0$. It obviously
results  that all form factors have the same ratio $0.35/0.8=0.44$ for their
$q^2=0$ value at $B$ versus $D$. The equality of these ratios is now in
agreement with what is expected, in the $m_{P_i}\gg m_f$ limit, from \hls and
the assumed \qq dependence: (\ref{hppz}).  But the value of the ratio, 0.44, is
larger than expected from the same relations. Although the latter should hold
only asymptotically, this suggests somewhat that the model violates \hls. This
can be proven rigourously in the same manner as above by noting that the
asymptotic behavior (\ref{nr}) at $q^2 = 0$ contradicts the relations
(\ref{hppz}).

The model gives a reasonable value $R =
1.61$, but $R_L = 0.36$ is too low, which seems to have escaped notice, with
the recent exception of \cite{gourdin}. Overall one could estimate that this
model is not faring too badly. This is however obtained by form factors (table
\ref{tab-RRL}) rather different from the form factors we advocate by appealing
to asymptotic principles (Soft-scaling-Pole solution in table \ref{tab-RRL}):
$A_2/A_1$ is sensibly higher, and $V/A_1$ is sensibly lower.

\subsubsection{Altomari Wolfenstein Model.}

We quote this model \cite{altomari} as an interesting proposal although it has
not been applied to the \dk and \bpsi phenomenology. In this model one assumes
that the non relativistic quark model is valid at \qmax, completed by a vector
dominance assumption for the \qq dependence. With such asumptions, it is easy
to see that \hls is satisfied asymptotically. On the other hand, it is obvious
that the model has not the \qq dependence required to satisfy \hhs away from
\qmax when both hadrons are made heavy.

\subsubsection{ISGW Quark Model.}

Although one is tempted to classify it among the non-relativistic models, it
results from
a modification of the NR model form factors which is no more the one predicted
by the wave functions; this has then some common ``hybrid'' spirit with the
previous A-W model. The justification given is however different: to cure  the
failure of the NR approximation,
an  ad hoc adjustment of the slope is made to take into
account relativistic effects which are indeed expected to enlarge the slope.

More precisely this adjustment consists in making in the NR formulae the
replacement:

\be \ved \to \frac {1} {\kappa^2} \frac
{m_f}{m_i}(q^2_{max}-q^2)\label{gisw}\ee
where is a $\kappa^2$ phenomenological factor $\simeq 0.5$. However we do not
think
that this prescription suffices to account for the variety of the expected
relativistic effects that will be discussed in section \ref{sec-oqm}.

One feature of the model is that in the transition to $0^-$ and $1^-$ all the
form factors are equal to their \qmax value times a {\it common}  exponential
function of \qq; the form factors are then easily found to respect exactly the
asymptotic \hl scaling; on the contrary they obviously violate the \hh scaling
except at \qmax, due to an inappropriate \qq dependence. This is bothering
since
the model should apply without any change to the \hh case. This failure is
easily
understandable since the relativistic boost of spin, which is
necessary to obtain the \hh scaling away from \qmax, is missing in the model.

The failure of the model for $\Gamma_L / \Gamma_{tot}$ corresponds to
the fact that $A_2/A_1$ is much too large (table \ref{tab-RRL}). This in turn
is
related to the fact that the form factor ratios $A_2/A_1$, already  too large
in \dk, is still increased up to \bpsi. Indeed, although this increase is
soften at \qmax by the light final masses, being  independent of \qq, this
ratio is not further depressed by a faster decrease of $A_2$ with respect to
$A_1$ as would be obtained by our (\ref{sp}) Ansatz.

Finally, it should be noticed that in some decays where the final state has a
large
velocity ($B\to \pi l\nu$) there is a dramatic suppression of the form factor
because of the exponential fall-off. We attribute this drawback to the absence
of Lorentz contraction of the wave function.

\subsubsection{Orsay Quark Model for form factors.}
\label{sec-oqm}

The Orsay Quark Model (OQM) for form factors is a semi-relativistic weak
binding
model that has been presented in the lectures \cite{gif-91} (see older
references therein) and that will be described in detail
elsewhere \cite{oqm}. In this section, $M, P, E$ refer to hadron masses,
energies momenta, while  $m, \vec p, \epsilon $ to quark masses, momenta,
energies.

 The model incorporates two main relativistic effects of {\it the center of
mass motion}: the Lorentz contraction of the wave
function and the Lorentz boost of the spinors. On the other hand, we adopt a
weak-binding treatment. {\it One makes everywhere the approximation of
retaining
only linear terms in internal momenta} and one sets $M_i$ and $M_f$ equal to
the sum of corresponding quark constituent masses. Therefore, as we explain in
detail below, we do not consider it as a truly phenomenological model; it is
rather an analytical instrument to discuss the specific effects of
center-of-mass motion.
We give now only the general principles behind it and write down the explicit
form
of the form factors.  The total wave function writes

\be \psi^{tot}_{\vec{P}}(\{ \vec{p}_i \}) = \delta \left ( \sum_i \vec{p}_i -
\vec{P} \right )
\psi_{\vec{P}} (\{ \vec{p}_i \}) \label{wf}\ee
where the internal wave function is given by

\be \psi_{\vec{P}} ( \{ \vec{p}_i \} ) = N \left [ \prod_i S_i(\vec{P}) \right
] \psi_{{\bf P}=0} (
\{ \tilde{\bf p}_i \} ) \label{iwf}\ee
with

\[ {\vec{\tilde p}}_{iT} \equiv \vec{p}_{iT} \]
\[ \tilde{p}_{iz} \equiv {E \over M} p_{iz} - {P \over M} \varepsilon_i  \]
\[ \tilde{\varepsilon}_i \equiv {E \over M} \varepsilon_i - {P \over M} p_{iz}
\simeq m_i \]
and where

\[ S_i(\vec{P}) = \sqrt{{E + M \over 2M}} \left ( 1 + {\vec{\alpha}_i.\vec{P}
\over E + M}
\right ) \]
is a the Lorentz boost acting on the spinors

\[ u_{\vec{P}=0} \simeq \left ( \vbox{\hbox{$\chi$}
\hbox{${\vec{\sigma}.\tilde{\bf
p} \over 2m}
\chi$ }} \right ) \]
the normalization being global for the internal wave function:

\[ \int \psi_{\vec{P}}^+(\{ \vec{p}_i \}) \psi_{\vec{P}}(\{ \vec{p}_i \})
\delta \left ( \sum_i
\vec{p}_i - \vec{P} \right ) \prod_i d\vec{p}_i = 2E \]

\noindent giving

\[ N = \sqrt{2M} \sqrt{1 - \beta^2} \ \ \ . \]

The matrix element of an operator acting on the quark 2 will read, in the equal
velocity frame (a
collinear frame where the velocities are equal in magnitude and opposite in
direction), after
some algebra :

\[ \int \psi_{\vec{P}_f}^+ ( \{ \vec{p}\ '_i \} ) O(2) \psi_{\vec{P}_i} (\{
\vec{p}_i \}) \delta \left
( \sum_i \vec{p}\ '_i - \vec{P}_f \right ) \delta \left ( \sum_i \vec{p}_i -
\vec{P}_i \right )
\prod_i d \vec{p}_i \ d \vec{p}\ '_i \delta \left ( \vec{p}_1 - \vec{p}\ '_i
\right ) \delta \left (
\vec{p}_2 - \vec{p}\ '_2 - \vec{q} \right ) \]

\[ = {N_i N_f \over \sqrt{1 - \beta^2}} \delta \left ( \vec{P}_i - \vec{P}_f -
\vec{q} \right ) \int
\psi_{\vec{P}_f = 0}^+ \left ( \vec{p}_1 - {m_1 \over M_f} \tilde{\bf P}_f,
\vec{p}_2 + {m_1 \over
M_f} \tilde{\bf P}_f \right ) S_2^+(\vec{P}_f) O(2) S_2(\vec{P}_i) \]

\be \psi_{\vec{P}_i=0} \left ( \vec{p}_1 - {m_1 \over M_i} \tilde{\bf P}_i ,
\vec{p}_2 + {m_1 \over
M_i} \tilde{\bf P}_i \right ) \delta \left ( \sum_i \vec{p}_i \right ) \prod_i
d \vec{p}_i \ \ \ . \label{matrel}\ee

The wave function at rest is assumed to be given by the
harmonic oscillator potential. \par

With these ingredients one can compute all the form factors we are interested
in. Calling $M_i, M_f$ the initial and final hadron masses, $m_i, m_f$ the
initial and final active quark masses, we find:

\be f_+(q^2) =  {\sqrt{4M_i M_f} \over M_i + M_f } {1 \over  1 - {q^2 \over
\left ( M_i + M_f \right )^2}  } I(q^2) \left [ 1 + {M_i - M_f \over M_i + M_f}
X \right ] \label{fplus}\ee

\be V(q^2) =  {\sqrt{4M_i M_f} \over  M_i + M_f } {1 \over  1 - {q^2 \over
\left ( M_i + M_f \right )^2} } I(q^2) (1 + Y) \label{v}\ee

\be A_1(q^2) =  {\sqrt{4M_i M_f} \over M_i + M_f } I(q^2) \left [ 1 + {  \left
(
M_f - M_i \right )^2 - q^2  \over \left ( M_f + M_i \right )^2 - q^2 } Y
\right ] \label{a1}\ee

\be A_2(q^2) =  {\sqrt{4M_i M_f} \over M_i + M_f} {1 \over  1 - {q^2 \over (M_i
+ M_f)^2}
 } I(q^2) \left [ 1 + { \left ( M_i^2 - M_f^2 \right ) - q^2  \over
 \left ( M_i + M_f \right )^2 - q^2 } Y - {2 M_f \left ( M_i + M_f \right )
\over
\left ( M_f + M_i \right )^2 - q^2 } X \right ] \label{a2}\ee
where, for the harmonic oscillator potential:

\be I(q^2) = \left ( {2 R_i R_f \over R_f^2 + R_i^2} \right )^{{3 \over 2}} exp
\left ( - {2m^2 R_i^2
R_f^2 \over  R_f^2 + R_i^2 }\left[ { \left ( M_f - M_i \right )^2 - q^2
\over \left ( M_f + M_i \right)^2 - q^2 } \right]\right ) \label{i}\ee

\noindent and

\be X = - {m \over R_f^2 + R_i^2} \left ( {R_f^2 \over m_i} - {R_i^2 \over m_f}
\right ) \qquad Y =
{m \over R_f^2 + R_i^2} \left ( {R_f^2 \over m_i} + {R_i^2 \over m_f} \right )
\label{xy}\ee

\noindent parametrize corrections to the scaling limit, proportional to the
spectator quark mass $m$
and $R_i$ and $R_f$ are the radii of the initial and final mesons. \par

\paragraph{ Isgur Wise scaling.}
It is important to realize that, with such a model, in the limit where
both $m_i$ and $m_f$ are made heavy, {\it one obtains exactly the whole set of
scaling relations of Isgur-Wise}, eq. (\ref{hh}). The scaling function
$\xi(v_i.v_f)$  depends of course on the potential except for the relation
$\xi(1)=1$. In the case of the harmonic oscillator potential:
\[	\xi(v_i.v_f) = {2 \over 1 + v_i.v_f} exp \left [ - {m^2R^2 \over \sqrt{2}}
\left (
{v_i.v_f - 1
\over v_i.v_f+1} \right ) \right ]. \]

\noindent The corresponding slope at the origin, within the weak binding and
linear approximation, is :

\[	\rho^2 = - \xi '(1) = {1 \over 2} + {m^2 R^2 \over 2\sqrt{2}}\sim 0.9 \]

\noindent where $R$ is the radius of a light-light meson.

\paragraph{\qq Dependence.}
The expressions for the form factors above show that the $q^2$ dependence of
$A_1(q^2)$ is very weak especially near $q^2=0$
and that the $q^2$ dependence of the form factors $f_+(q^2)$, $V(q^2)$ and
$A_2(q^2)$ is dominated by
the same kinematic pole that appears in the Isgur-Wise relations in the
heavy-heavy case (3). In the
model this kinematic pole comes simply from the Lorentz factor

\[ 1 - \beta^2 = {4M_iM_f \over \left ( M_i + M_f \right )^2} {1 \over  1-{q^2
\over \left ( M_i + M_f \right )^2} } \]

\noindent that does not affect $A_1(q^2)$ because this form factor is related
to a purely transverse
component of the axial current. The  \qq dependence of $A_1$ comes essentially
from the exponential and it becomes rather weak when $\ved$ is large, i.e.
near $q^2=0$; indeed it tends to a constant. This would be true for any
potential.
It is a simple effect of the Lorentz contraction.\par

This model's prediction is therefore similar to the QMI Ansatz, eqs. (\ref{sp})
and (\ref{one}). The $q^2$ dependence of form factor ratios is almost the same
in both models.

\paragraph{Corrections to scaling.}

For finite $m_i, m_f$, the \hhs laws are broken in OQM by various effects

i) the radii in $I(q^2)$ depend on the flavour

ii) the terms containing X and Y are of order $m/m_i,_f$. The scaling is broken
because the spectator mass is no longer negligible.

On the whole, these scaling violations are rather small except for $f_+$ and
especially for $V$. That the latter is large is easily understandable even in
the
most naive non relativistic approximation. Indeed, the $1+Y$ coefficient in
(\ref{v}) varies from 2 for $m_i=m_f=m$ to 1 when $m_i, m_f \gg m$.

\paragraph{Phenomenological shortcomings.}

Although the model is unique in obeying the full set of asymptotic scaling laws
in contrast to previous models, it is not a satisfactory phenomenological
model, because it lacks essential effects. First, due to the weak-binding
approximation, the axial to vector matrix elements at \qmax are reduced to
their static SU(6) value. For instance, one finds in this model for the nucleon
axial to vector  current coupling ratio: $G_A/G_V=5/3$ (sometimes written
$-g_1/f_1$) \cite{lopr}, which is too large by $\sim \sqrt{2}$. The same thing
happens in $D \to K^\ast l\nu$ and could also explain why $\Gamma(D\to K^\ast
l\nu)/\Gamma(D\to K l\nu)$ is predicted too large by a factor 2 in most quark
models. Indeed, the axial form factors $A_1, A_2$ are found too large with
respect to the vector ones. In the OQM:

\[ A^{sc}_1(0)=0.89,\quad A^{sc}_2(0)=0.87,\quad V^{sc}(0)=1.15,\quad
f^{sc}_+(0)=0.74\]

\be \frac{\Gamma_L}{\Gamma_{tot}}=0.41\qquad \frac{\Gamma(B\to
K^\ast)}{\Gamma(B\to K)}=4.75 \label{numbers}\ee

This is quite bad. To cure the problem with $G_A/G_V=5/3$, we have adopted in
the past the old recipe of multiplying the axial current by an adhoc factor
$g_A=.7$  \cite{kokkedee}. This gives, multiplying all axial form factors by
$g_A$:

\be \frac{\Gamma_L}{\Gamma_{tot}}=0.34\qquad \frac{\Gamma(B\to
K^\ast)}{\Gamma(B\to K)}=2.8 \label{numbers2}\ee

 Of course this is still unsatisfactory. Besides, such a recipe has no
theoretical grounding. One should include systematically the binding
corrections, which are known to correct the discrepancy for $G_A/G_V$
\cite{bogoliubov}, \cite{gavela}.

Second, there is another difficulty concerning $f_+$. It tends to be too
small by itself at large recoil for the pion final state and this cannot be
easily explained by
binding corrections which should have a reductive effect. This smallness is
connected with the small mass of the $\pi$ as observed in \cite{gif-91}.
We suspect that this could be connected with the Golsdtone character of the
$0^-$. Indeed, in the approximations of the present Quark Model the $0^-$ and
$1^-$ should be degenerate.

\subsection{The phenomenological analysis by Gourdin, Kamal and Pham.}

While we were in the process of writing this paper we received a paper by
Gourdin, Kamal
and Pham \cite{gourdin} which also study the relation between \bpsi and \dk
experiments with the help of \hl scaling rules, and also confront some
theoretical approaches with \bpsi experiments. We agree with these authors on
their main conclusion that the current models do not fit the  \bpsi
data. We differ with them in two respects. They have used all over the pole
dominance for the \qq dependence of all form factors, while we conclude to a
different \qq  behaviour of the different form factors.  Second, their eq.
(29) which they present as {\it the} \hl scaling relation is in our opinion
{\it one possible} form of the $m_Q$ dependence. The detailed dependence on
masses contained in these eqs. is an arbitrary Ansatz, though an admissible
one, since it
is compatible with the asymptotic scaling relations. This Ansatz contains
non-asymptotic corrections which amount to a softening of the asymptotic
scaling law. Therefore we do not the least object to its use as a model, but
only to
the claim that it should be taken as {\it the} \hl Isgur Wise scaling law.
Notice that their eq. (29) differs from the mass dependence in our Ansatz
(\ref{sp}) and (\ref{one}).
Finally we agree with the claim by these authors that \dk and \bpsi data are
difficult to reconcile within the \hls laws. This difficulty, beyond the  \glgt
problem stressed by the authors, shows up in $\Gamma(K^\ast)/\Gamma(K)=2.86$
predicted from their eq. (33).

\section{Conclusions.}
\label{sec-conc}

All the approaches we have considered in this paper encounter difficulties in
accounting for the \bpsi data, particularly with the large \glgt (CLEO and
ARGUS data). At present it seems safer to keep  open three possibilities to get
out of this problem.
\begin{itemize}

\item Experiment may not have yet delivered its ultimate word, as the variation
between different experiments seem to indicate, and it might evolve towards
data easier to account for.

\item Although we did not discuss the factorization assumption, it should be
kept in mind that it rests on no theoretical ground for color suppressed decay
channels, as is the case for \bpsi.

\item Finally, models may be wrong. This will now be discussed in more details.
\end{itemize}

Of course, the first requirement for any model is to fulfill the \hls
relations. This has been seen not to be the case for the most popular BSW I and
BSW II models, notwithstanding their relatively good empirical successes.

Our analysis has allowed to extricate from data some general trends, namely
``softened'' scaling, a sensibly different \qq behaviour of $A_1$ versus $A_2,
V, f_+$, and $A_1$ slowly varying with \qq. Ans\"atze that take these
indications as a guide, obtain better values for \glgt, and a more reasonable
$a_2$, although
there remains a general tendency to underestimate \glgt with respect to present
data.
Let us now comment on these general trends.

Data definitely exclude ``hard scaling'' i.e. the strict application of
asymptotic \hls formulae in the finite mass domain. We have proposed a
``softened'' Ansatz which is based on an extension of \hhs relations down to
the light final meson case, with some rescaling. In fact this is equivalent to
assuming a precocious scaling for the axial and vector current {\it matrix
elements}. Consequently, the ratio $A_2/A_1$ does not increase too fast with
the heavy mass.

There are indications from lattice calculations, form Quark Model, and to some
degree from phenomenology, that $V$ should undergo an even softer scaling.

Another consequence of the above Ansatz, as well as of the Orsay Quark Model is
that $A_2/A_1, V/A_1$ and $f_+/A_1$ should have a pole like behaviour in \qq,
leading to an increase with \qq. This improves the agreement with \bpsi data,
and seems to be corroborated by QCD Sum Rules calculations.

$D\to Kl\nu$ experiments seem to show a pole like behaviour for $f_+(q^2)$.
Combined with our preceding Ansatz for the ratios, this implies an
approximately constant $A_1(q^2)$. This particular \qq behaviour is
corroborated
by the Orsay Quark Model, while QCD Sum Rules give \qq dependence of $A_1$ that
never increases very fast, although different detailed shapes are proposed.
Lattice calculations, within large errors, might give the same indication.

\section*{Acknowledgements.}
We acknowledge  Nathan Isgur,  Stephan Narison and Daryl Scora for interesting
comments and useful informations.
We are particularly indebted to Patricia Ball and to the European Lattice
Collaboration for providing us with their results concerning semi-leptonic form
factors, and to  Asmaa Abada, Philippe Boucaud and Jean Pierre Leroy for
enlightening
discussions.
This work was supported in part by the CEC Science Project SC1-CT91-0729 and by
the Human Capital
and Mobility Programme, contract CHRX-CT93-0132.

\end{document}